\documentclass[
    twocolumn,
    superscriptaddress,
    showpacs,
    preprintnumbers,
    amsmath,
    amssymb,
    nofootinbib
]{revtex4-2}
\usepackage{graphicx}
\usepackage{dcolumn}
\usepackage{bm}
\usepackage{tabularx}
\usepackage[normalem]{ulem} 
\newcolumntype{M}{>{\centering\arraybackslash}m{1.85cm}}
\usepackage[export]{adjustbox}
\usepackage{float}
\usepackage{placeins}
\usepackage{braket}
\usepackage{lipsum}
\usepackage{amsmath}
\usepackage{tikz}
\usetikzlibrary{quantikz2}
\tikzset{qgrp/.style={dashed,rounded corners,inner xsep=1pt,inner ysep=1pt}}
\usepackage{multirow}
\newcommand{\expval}[1]{\left\langle#1\right\rangle}
\newcommand{\PREP}{\mathrm{PREP}}
\newcommand{\SELECT}{\mathrm{SELECT}}
\newcommand{\Real}{\operatorname{Re}}

\usetikzlibrary{shapes.geometric, arrows.meta, positioning, calc}
\graphicspath{{figure/}}
\usepackage{xcolor}   
\usepackage{mathrsfs}

\usepackage{caption}
\usepackage{subcaption}

\usepackage{hyperref}
\hypersetup{
    colorlinks=true, 
    linktoc=all,     
    urlcolor  = cyan,
    citecolor = blue,
    linkcolor=blue,  
}

\makeatletter
\newcommand{\colorcaption}[2][]{%
  \begingroup%
  \renewcommand{\@caption@fignum@sep}{ (Color online). }%
  \caption[#1]{#2}%
  \endgroup%
}
\makeatother
\newcommand{\orcid}[1]{\href{https://orcid.org/#1}{\hskip2pt\includegraphics[width=9pt]{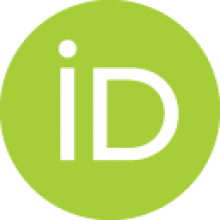}}}
\begin{document}
\title{Quantum computation of nuclear observables using LCU techniques}

\author{Rohit M. Shinde}
\email{rohit\_ms@ph.iitr.ac.in}
\affiliation{Department of Physics, Indian Institute of Technology Roorkee, Roorkee 247667, India}

\author{Chandan Sarma}
\email{c.sarma@surrey.ac.uk}
\affiliation{School of Mathematics and Physics, University of Surrey, Guildford, GU2 7XH, Surrey, United Kingdom}

\author{Praveen C. Srivastava}
\email{praveen.srivastava@ph.iitr.ac.in}
\affiliation{Department of Physics, Indian Institute of Technology Roorkee, Roorkee 247667, India}

\date{\hfill \today}

\begin{abstract}
We present a quantum computing framework for calculating the reduced matrix elements of the electric quadrupole ($\mathcal{M}_{E2}$) and magnetic dipole ($\mathcal{M}_{M1}$) moments within the nuclear shell model. The operators are expanded as weighted sums of Pauli strings, and their expectation values are obtained using the Hadamard test of linear combination of unitaries (HoLCU); this method combines the linear combination of unitaries (LCU) with a Hadamard test. HoLCU requires a single circuit and the measurement of a single qubit, without post-selection on the ancilla register. The state-preparation part of the LCU uses the Grover--Rudolph algorithm, decomposed with the M\"ott\"onen decomposition, and the selection part uses unary iteration, so that all multi-controlled gates are replaced by 1- and 2-qubit gates. This comes at the cost of a few auxiliary qubits, but it reduces the circuit depth and the number of 2-qubit gates by up to two orders of magnitude. We use this method to evaluate the reduced matrix elements for two lighter two-nucleon systems: $^6$He and $^{18}$O. We further use another method, HoLCU-div, primarily for the noisy simulations of $^{18}$O, in which the circuit size is reduced by grouping Pauli strings with the same weights, but it requires running multiple quantum circuits. We apply these methods to the first excited state ($2^+$) of these two nuclei, obtained by following a variational quantum eigensolver (VQE) method. After initial testing with a noiseless simulator, the circuits were executed on a noisy simulator (\texttt{FakeFez}) followed by error-mitigation techniques including zero-noise extrapolation (ZNE) and readout-error mitigation (REM). Finally, we executed the $^6$He circuits for both observables on IBM hardware (\textit{ibm\_fez}) followed by error mitigation techniques.

\end{abstract}


\maketitle

\section{INTRODUCTION}

Quantum computing algorithms and frameworks for solving various problems in quantum many-body systems have advanced rapidly over the last decade~\cite{chuang, nisq, QC_n1,Lacroix_PRL,James1,Sarma_2026_PLB,James, Ayral_2023,Lee, GarciaRamos_2025, Klco_2022, Yoshida}. This has especially motivated efforts to map nuclear theory calculations onto quantum computers, as we reach computational bottlenecks when working with larger Hilbert spaces. Methods such as the variational quantum eigensolver (VQE) and ADAPT-VQE have been used to calculate the ground-state energies of nuclear shell-model Hamiltonians~\cite{Perez2_2023, Chandan_2023, Bharti_Bhoy, QC_n3, QC_n4, QC_n5, Denis_2024, Zhang_Lacroix_2025, Robin_2023, Romero_2022}. Progress has also been made in calculating the low-lying excited states of various nuclei using the variational quantum deflation (VQD) method, the subspace-search variational quantum eigensolver (SSVQE), and the quantum generator coordinate method (QuGCM)~\cite{Higgott_2019, Jiang_2023, Nakanishi_2019, Maheshwari_2026, Zheng_2023, Pandey_2026}.
While many of the above-mentioned algorithms are usually based on the Jordan-Wigner (JW) encoding~\cite{Jordan_Wigner_1928}, there are some efforts to use alternative encodings, including Gray-code encoding~\cite{graycode, Siwach_2021, Singh_2025} and other problem-specific encodings~\cite{Costa_2026, Sarma_2026, Yoshida_2026, Sarma_2026_PLB, Liu_2025}.

However, these works have focused mainly on ground- and excited-state energies, while the evaluation of electromagnetic observables, such as magnetic dipole and electric quadrupole moments, directly on the quantum computer has received comparatively little attention. This motivates us to construct quantum circuits that compute these electromagnetic observables on the quantum computer itself, and to build a pipeline that yields both the energies and the electromagnetic observables of the nuclear systems under study.

Linear combinations of unitaries (LCU) have been used in nuclear physics in several settings~\cite{Childs_Wiebe_2012, Berry_2015}. In the first, the LCU formalism is used to solve the shell-model Hamiltonian itself, as in the QCSH package~\cite{Lv_2023}, which computes binding energies of light nuclei entirely on the quantum computer. In the second group, one work uses LCU to apply non-unitary excitation operators in the calculation of the $n(p,d)\gamma$ reaction~\cite{Roggero_2020}, while another simulates an $M1$ transition between the ground and first excited states of the triton, with both states prepared using variational methods~\cite{Nigro_2025}. In these works, the operator is applied probabilistically, so a run is accepted only when an ancilla register is measured in a prescribed state. In the third work, LCU is used to evaluate nuclear observables, such as the quadrupole moment of the deuteron~\cite{pooja2}. The expectation value is obtained from Hadamard tests on the individual unitaries and from SWAP and D-SWAP tests that require a second copy of the register.

In this work, we evaluate the expectation value $\langle\psi|A|\psi\rangle$ of the reduced matrix elements of the electric quadrupole and magnetic dipole moment operators of the nuclear shell model on a quantum computer. These operators can be written as a linear combination of unitaries (LCU)~\cite{Childs_Wiebe_2012, Berry_2015}, $A=\sum_k c_kU_k$, where each $U_k$ is a Pauli string and $c_k$ are their corresponding coefficients. The standard LCU procedure applies $A$ to a target state through two operations: $\mathrm{PREP}$, which encodes the magnitudes of the coefficients $c_k$ in the amplitudes of an ancilla register, and $\mathrm{SELECT}$, which applies each unitary $U_k$ to the target state, controlled by the state of the ancilla register. Its drawback is that it is probabilistic, since the operator is only applied when the ancilla register is found in a $\ket{0}^{\otimes m}$ state. We remove this limitation by combining the LCU with the Hadamard test (HoLCU)~\cite{MataAli_2025}. This results in a deterministic procedure in which the expectation value is obtained by measuring only the Hadamard qubit $\ket{0}_H$, with no post-selection.

\begin{figure*}[t]
    \centering
        \includegraphics[width=0.98\textwidth]{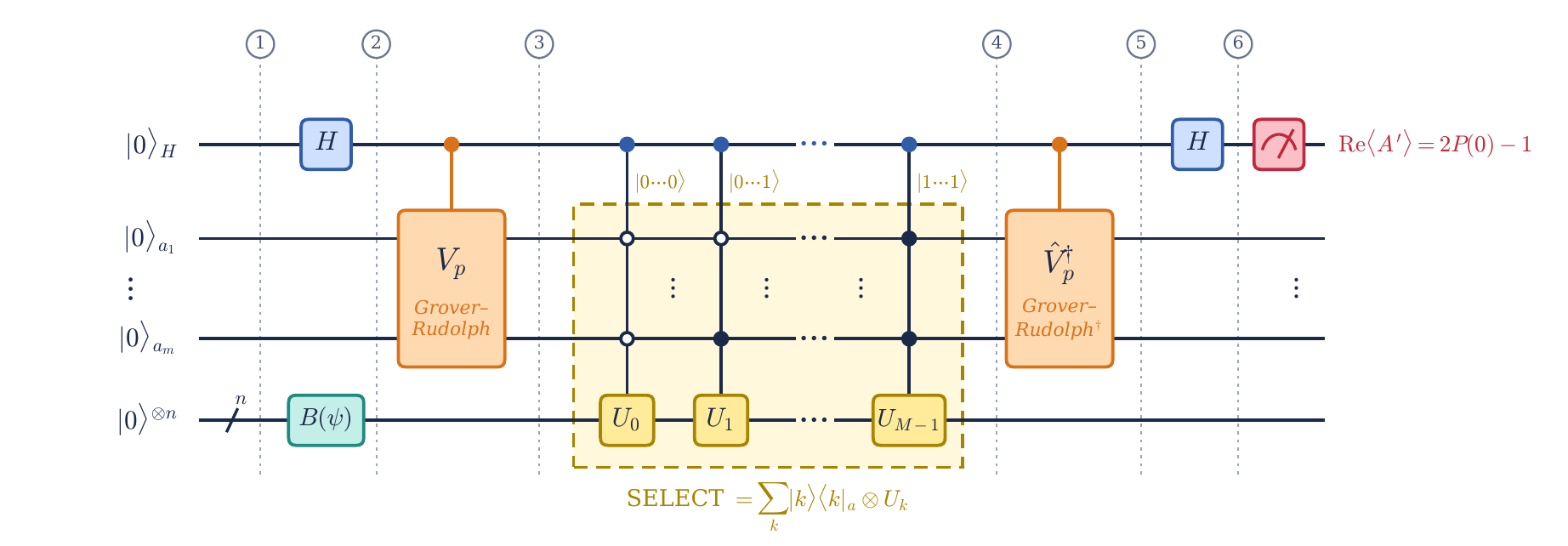}
    \caption{\small Gate-level circuit of the HoLCU for estimating the expectation value $\langle A\rangle$ of an operator $A=\sum_{k=0}^{M-1}c_k U_k$ with respect to the state $\ket{\psi}$. The $\mathrm{PREP}$ block (orange), $V_p$, and its inverse $\hat{V}_p^{\dagger}$ are constructed with the Grover-Rudolph algorithm. The $\mathrm{SELECT}$ block (yellow) applies $U_k$ conditioned on the ancilla state $|k\rangle$; only the term $U_0$, associated with $|0\cdots0\rangle_a$, needs to be controlled by the Hadamard qubit. The measurement is performed only on the Hadamard qubit, and $\mathrm{Re}\langle A\rangle = 2P(0)-1$.}

    \label{fig:HoLCU_diagram}
\end{figure*}

The resulting circuit has three registers: the Hadamard qubit register, the LCU ancilla register, and the register that holds the nuclear wave function as shown in Fig.~\ref{fig:HoLCU_diagram}. Its main building blocks are compiled as follows. The $\mathrm{PREP}$ operation is implemented with the Grover--Rudolph algorithm~\cite{Grover_Rudolph_2002, Ramacciotti_2024, Sun_2023}, whose rotations are decomposed into elementary gates with the M\"ott\"onen decomposition~\cite{Mottonen_2004, Jaunin_2025}. The $\mathrm{SELECT}$ operation is implemented with unary iteration~\cite{Babbush_2018, Wierichs_2025}, which avoids multi-controlled gates for each term. To limit the circuit depth, which makes the results dominated by noise, we also implement a variant of HoLCU, called HoLCU-Div, at the cost of running multiple circuits.

This paper is arranged as follows. Section \ref{sec:framework} has many theoretical components which are organized in the following way: Section~\ref{NSM} introduces the notation of the nuclear shell model and the calculation of the observables, Section~\ref{VQE} describes how the wave function is prepared on a quantum computer using the variational quantum eigensolver (VQE), Section~\ref{HoLCU} explains how HoLCU operates, and Sections~\ref{GR-Algo} and~\ref{SELECT} then describe how $\mathrm{PREP}$ and $\mathrm{SELECT}$ are compiled into elementary gates. In Section \ref{results}, we have presented results for the nuclear observables of $^6$He and $^{18}$O using ideal and noisy simulators, along with zero-noise extrapolation and readout-error mitigation. We finally conclude our work in Section \ref{summary}. Pauli-string decompositions of the operators and quantum circuits, as well as their decompositions, can be found in the 
Appendices~\ref{app:appendixa}-\ref{app:appendixc}.

\section{Theoretical Framework}\label{sec:framework}

\subsection{Nuclear Observables in Shell Model framework} \label{NSM}

In the second quantization, the shell model Hamiltonian can be written as 
\begin{equation}
\label{eq1}
H = \sum_{i}\epsilon_i \hat{a}^\dagger_i  \hat{a}_i + \frac{1}{2} \sum_{i, j, k, l} V_{ijlk} \hat{a}^\dagger_i \hat{a}^\dagger_j \hat{a}_l \hat{a}_k.
\end{equation}
Here, $\hat{a}^\dagger_i$ and $\hat{a}_i$ are the creation and annihilation operators of a nucleon in state $|i\rangle$.
The single-particle energies and two-body matrix elements can be represented by  $\epsilon_i$ and $V_{ijlk}$, respectively.

Once we have solved the nuclear shell model Hamiltonian and obtained the wave function corresponding to our desired state, we can calculate the density matrix with respect to the operator, given by

\begin{equation}
\begin{split}
\rho^{\lambda}_{ab} ={}&
\sum_{m_a,m_b}
(-1)^{j_b-m_b}
\begin{pmatrix}
j_a & j_b & \lambda \\
m_a & -m_b & 0
\end{pmatrix}
\begin{pmatrix}
J_i & \lambda & J_f \\
J_z & 0 & J_z
\end{pmatrix}^{-1} \\
&\times
\sqrt{\frac{2J_f+1}{2\lambda+1}}\,
\left\langle
a^\dagger_{m_a}a_{m_b}
\right\rangle .
\end{split}
\end{equation}

The reduced matrix element of a one-body operator is obtained by 

\begin{equation}
(\psi_f J_f\|\mathcal{M}_{\sigma\lambda}\|\psi_i J_i)
=
\sum_{ab}
(a\|\mathcal{M}_{\sigma\lambda}\|b)\,
\rho^{\lambda}_{ab},
\end{equation}
where a, b are labels for single-particle orbits, and $(a\|\mathcal{M}_{\sigma\lambda}\|b)$ are the reduced one-body matrix elements for the operator $\mathcal{M}_{\sigma\lambda}$~\cite{Suhonen_2007, Johnson_2018}. For electric and magnetic transitions, $\sigma=E, M$, respectively. The $\lambda$ dictates the angular momentum corresponding to the operator.

\subsection{Wave function preparation using VQE} \label{VQE}
We use the variational quantum eigensolver (VQE) to compute the lowest eigenvalue of the shell-model Hamiltonian in the subspace of fixed total magnetic projection $M=2$. For both nuclei, the lowest state in this subspace is the first excited state, the $2^+$ state. VQE is based on the variational principle: the trial wave function is parameterized by an ansatz $\ket{\psi(\theta)}$ prepared on the quantum device, and the parameters $\theta$ are minimized with a classical optimizer. This makes VQE a hybrid quantum-classical algorithm, producing a shallower circuit and therefore well-suited for Noisy Intermediate-Scale Quantum (NISQ) devices.

For both nuclei, we use a problem-inspired ansatz. The $^6$He ansatz consists of a single single-excitation gate, and the $^{18}$O ansatz consists of eight double-excitation gates. Working at fixed $M=2$ simplifies the problem, because the single-particle states that have zero occupation probability in this sector can be dropped from the wave function register. As a result, the register needs only 3 qubits for $^6$He and 9 qubits for $^{18}$O. The remaining single-particle states are mapped onto these qubits with the Jordan--Wigner mapping. We have used  the CKPOT interaction~\cite{Cohen_Kurath_1965} for the $p$-shell nucleus $^6$He and the USDB interaction~\cite{Richter_2008} for the $sd$-shell nucleus $^{18}$O.

\subsection{Hadamard of Linear Combination of Unitaries} \label{HoLCU}
For the LCU part of the circuit, the operator is defined as 
\begin{equation}
 A=\sum_{k=0}^{M-1} c_k U_k,    
\end{equation}
where $U_k$ are unitaries and $c_k$ are the coefficients. Since we require a set of normalized coefficients, we need to rewrite our coefficients $c_k$ as
\begin{equation}
  \qquad \alpha_k=\frac{|c_k|}{N},
\end{equation}
 where $N=\sum_{k}|c_k|$, and the operator is rewritten as $A' = \frac{A}{N}$. The LCU protocol uses an ancilla register of $m=\lceil\log_2 M\rceil$ qubits and two operators~\cite{Childs_Wiebe_2012}:
\begin{align}
  \PREP\,\ket{0}^{\otimes m} &= \sum_{k=0}^{M-1}\sqrt{\alpha_k}\,\ket{k}, \label{eq:prep}\\
  \SELECT &= \sum_{k=0}^{M-1}\ket{k}\!\bra{k}\otimes \xi_k U_k , \label{eq:select}
\end{align}
where the phases $\xi_k$ have been absorbed into $\SELECT$. Applying $(\PREP^\dagger\otimes I)\,\SELECT\,(\PREP\otimes I)$ to $\ket{0}^{\otimes m}\ket{\psi}$ gives
\begin{equation}
  \ket{0}^{\otimes m}A'\ket{\psi}+\ket{\perp}A'_\perp\ket{\psi},
\end{equation}
where $\ket{\perp}$ denotes states of the ancilla register orthogonal to $\ket{0}^{\otimes m}$ and $A'_\perp$ is the complementary operator. In other words, the circuit is a block encoding of $A'$. Applying $A'$ is therefore probabilistic: it succeeds only when the ancilla register is found in $\ket{0}^{\otimes m}$.

After establishing the circuit of LCU, we can conduct the Hadamard test on it. The final state is given by
\begin{equation}
\begin{split}
  \tfrac12\ket{0,0}(I+A')\ket{\psi}+\tfrac12\ket{0,\perp}A'_\perp\ket{\psi}+\\
  \tfrac12\ket{1,0}(I-A')\ket{\psi}-\tfrac12\ket{1,\perp}A'_\perp\ket{\psi},    
\end{split}
\end{equation}
where the state is of the form $\ket{x}_H\ket{y}_a^{\otimes m}\ket{z}$, where $\ket{x}_H$ denotes the Hadamard qubit state, $\ket{y}_a^{\otimes m}$ denotes the LCU ancilla state, and $\ket{z}$ denotes the wave function as described in Ref.~\cite{MataAli_2025}. The probability of measuring the Hadamard qubit in $\ket{0}$ and $\ket{1}$ is
\begin{equation}
\begin{split}
P(0) ={}& \tfrac{1}{4}\left(1+2\Real\expval{A'}
+\expval{A'^\dagger A'}
+\expval{A{'}_\perp^\dagger A{'}_\perp}\right) 
\end{split}
\end{equation}
\begin{equation}
\begin{split}
P(1) ={}& \tfrac{1}{4}\left(1-2\Real\expval{A'}
+\expval{A'^\dagger A'}
+\expval{A{'}_\perp^\dagger A{'}_\perp}\right). 
\end{split}
\end{equation}

The final result is the same as that of the Hadamard test,
\begin{equation}
  \Real\expval{A}=N\,\Real\expval{A'}=N\,\big(2P(0)-1\big).
  \label{eq:holcu}
\end{equation}

For the case of HoLCU-div, when several terms of $A$ share the same coefficient, the state preparation can be simplified considerably. In this case we group the terms by $|c_k|$ into $M_t$ groups,
\begin{equation}
  A=\sum_{g=1}^{M_t}|c_g|\,B_g,\qquad B_g=\sum_{k\in G_g}\xi_k\,U_k ,
\end{equation}
where $G_g$ contains the $M_g$ Pauli strings with common magnitude $|c_g|$, and the signs $\xi_k=\pm1$ are kept inside $B_g$. Within a group, all the weights on the LCU ancilla registers are equal, so $\mathrm{PREP}$ is a uniform superposition over $2^{m_g}$ indices, with $m_g=\lceil\log_2 M_g\rceil$. It requires only Hadamard gates, controlled by the Hadamard qubit

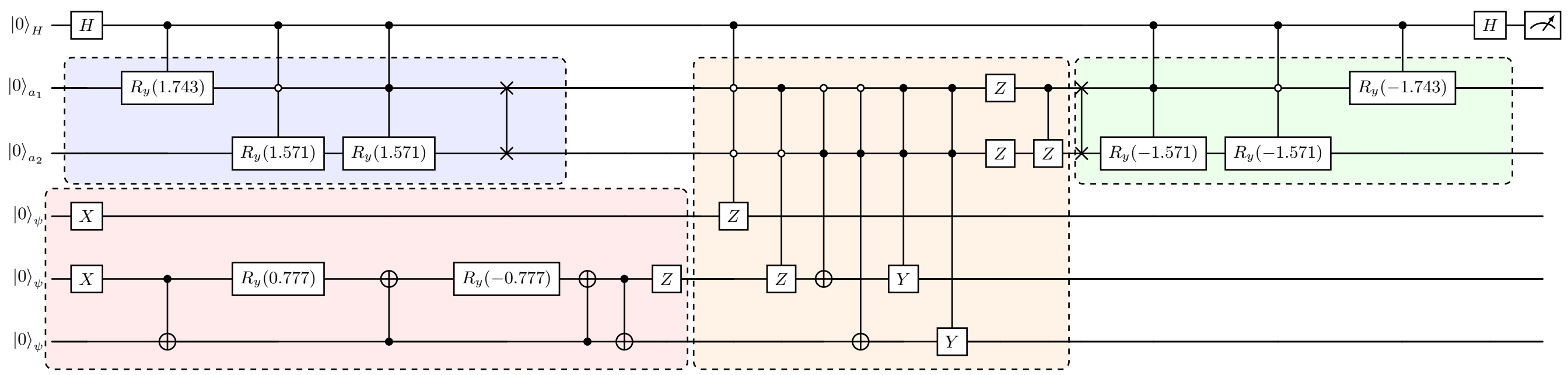
\begin{figure*}[t]
  \centering
  \resizebox{0.99\textwidth}{!}{%
\begin{quantikz}[row sep=0.6cm, column sep=0.35cm]
  \lstick{$\ket{0}_H$} & \gate{H}
    & \ctrl{1} & \ctrl{1} & \ctrl{1} & \qw & \qw & \qw & \qw & \qw
    & \ctrl{1}
    & \qw & \qw & \qw & \qw & \qw
    & \qw & \qw
    & \qw
    & \ctrl{1} & \ctrl{1} & \ctrl{1} & \gate{H} & \meter{} \\
  \lstick{$\ket{0}_{a_1}$} & \qw
  \gategroup[wires=2,steps=5,style={dashed,rounded corners,fill=blue!8,inner xsep=0.1pt},background,label style={label position=above,anchor=south,yshift=0.2cm}]{}
  & \gate{R_y(1.743)} & \octrl{1} & \ctrl{1} & \swap{1} & \qw & \qw & \qw & \qw
  \gategroup[wires=5,steps=9,style={dashed,rounded corners,fill=orange!10,inner xsep=0.1pt},background,label style={label position=above,anchor=south,yshift=0.2cm}]{}
    & \octrl{1} & \ctrl{1} & \octrl{1} & \octrl{1} & \ctrl{1} & \ctrl{1}
    & \gate{Z} & \ctrl{1}
    & \swap{1}
    \gategroup[wires=2,steps=5,style={dashed,rounded corners,fill=green!8,inner xsep=0.1pt},background,label style={label position=above,anchor=south,yshift=0.2cm}]{}
    & \ctrl{1} & \octrl{1} & \gate{R_y(-1.743)} & \qw & \qw \\
  \lstick{$\ket{0}_{a_2}$} & \qw & \qw & \gate{R_y(1.571)} & \gate{R_y(1.571)} & \targX{} & \qw & \qw & \qw & \qw
    & \octrl{1} & \octrl{2} & \ctrl{2} & \ctrl{3} & \ctrl{2} & \ctrl{3}
    & \gate{Z} & \gate{Z}
    & \targX{} & \gate{R_y(-1.571)} & \gate{R_y(-1.571)} & \qw & \qw & \qw \\
  \lstick{$\ket{0}_{\psi}$}
    \gategroup[wires=3,steps=9,style={dashed,rounded corners,fill=red!8,inner xsep=0.1pt},background,label style={label position=above,anchor=south,yshift=0.2cm}]{}
    & \gate{X} & \qw & \qw & \qw & \qw & \qw & \qw & \qw & \qw
    & \gate{Z} & \qw & \qw & \qw & \qw & \qw
    & \qw & \qw
    & \qw & \qw & \qw & \qw & \qw & \qw \\
  \lstick{$\ket{0}_{\psi}$} & \gate{X} & \ctrl{1} & \gate{R_y(0.777)} & \targ{} & \gate{R_y(-0.777)} & \targ{} & \ctrl{1} & \gate{Z} & \qw
    & \qw & \gate{Z} & \targ{} & \qw & \gate{Y} & \qw
    & \qw & \qw
    & \qw & \qw & \qw & \qw & \qw & \qw \\
  \lstick{$\ket{0}_{\psi}$} & \qw & \targ{} & \qw & \ctrl{-1} & \qw & \ctrl{-1} & \targ{} & \qw
    & \qw & \qw & \qw & \qw & \targ{} & \qw & \gate{Y}
    & \qw & \qw
    & \qw & \qw & \qw & \qw & \qw & \qw
  \end{quantikz}%

  }
  \caption{Raw HoLCU circuit for the calculation of $\mathcal{M}_{E2}$ for the $2^+_1$ state of $^6$He. The qubit $\ket{0}_H$ is the Hadamard qubit, $\ket{0}_{a_1}$ and $\ket{0}_{a_2}$ are the LCU ancilla qubits, and the three $\ket{0}_{\psi}$ qubits encode the wave function. The blue dashed box marks the $\mathrm{PREP}$ part, the orange dashed box the $\mathrm{SELECT}$ part, and the green dashed box the inverse of the $\mathrm{PREP}$ part, $\mathrm{PREP}^{\dagger}$. The red region prepares the wave function, with parameters obtained from noiseless VQE simulations. The measurement is performed only on the Hadamard qubit.}
  \label{fig:Raw_6He}
\end{figure*}
$\ket{0}_H$. The signs are applied in $\mathrm{SELECT}$ by a $Z$ gate on the last auxiliary qubit introduced by the unary iteration method. If $M_g$ is not a power of two, the group is padded with $p_g=2^{m_g}-M_g$ identity operators, which are left as empty slots in $\mathrm{SELECT}$.

Each group is then evaluated with its own HoLCU circuit. Since equal weights of $2^{-m_g}$ are assigned to each ancilla qubit, the probability $P_g(0)$ of measuring the Hadamard qubit in $\ket{0}$ gives
\begin{equation}
\begin{aligned}
  \operatorname{Re}\langle B_g\rangle
  &=2^{m_g}\bigl(2P_g(0)-1\bigr)-p_g,\\
  \operatorname{Re}\langle A\rangle
  &=\sum_{g=1}^{M_t}|c_g|\operatorname{Re}\langle B_g\rangle .
\end{aligned}
\end{equation}

The padding term $p_g$ removes the contribution $\langle I\rangle=1$ of each padded identity. The method uses $M_t$ circuits instead of one, but each circuit has fewer ancilla qubits and a shallower $\mathrm{PREP}$. The method is advantageous when the coefficients are strongly degenerate, $M_t\ll M$.

\subsection{State Preparation algorithms} \label{GR-Algo}
The $\PREP$ operator of Eq.~\eqref{eq:prep} loads the amplitudes $\sqrt{\alpha_k}$ into the ancilla register. Since they are real and non-negative, no relative phases need to be prepared, reducing the problem to amplitude embedding using the $R_y$ rotations.

\paragraph{Grover--Rudolph.} The algorithm~\cite{Grover_Rudolph_2002} builds the state by successive bisections. In step $j=1,\dots,m$, the basis states are grouped into $2^{j-1}$ blocks, one for each value of the first $j-1$ qubits. Each block is split into a left half, where qubit $j$ is $\ket{0}$, and a right half is $\ket{1}$. The rotation on qubit $j$ is controlled by the first $j-1$ qubits, so each block $l\in\{0,\dots,2^{j-1}-1\}$ receives its own angle:
\begin{equation}
  \theta^{(j)}_l=2\arccos\sqrt{\frac{\sum_{k\in\text{left half of block }l}\alpha_k}{\sum_{k\in\text{block }l}\alpha_k}},
  \label{eq:gr-angles}
\end{equation}
with $\theta^{(j)}_l=0$ if the denominator vanishes. The fraction inside the root is the probability of finding qubit $j$ in $\ket{0}$ given that the register is in block $l$. Because $R_y(\theta)\ket{0}=\cos\frac{\theta}{2}\ket{0}+\sin\frac{\theta}{2}\ket{1}$, this probability equals $\cos^2(\theta/2)$, and Eq.~\eqref{eq:gr-angles} is the solution of that condition.  After $m$ steps, the exact basis state $\ket{k}$ is reached along with the corresponding coefficients $\alpha_k$ and the final state is the desired $\mathrm{PREP}$ state, $\sum_k\sqrt{\alpha_k}\ket{k}$. Step $j$ contains $2^{j-1}$ rotations, each with $j-1$ controls, so the number of multi-controlled gates grows exponentially with $j$. The exact preparation of a general state necessarily requires exponentially many gates.


\paragraph{M\"ott\"onen decomposition.} The multi-controlled rotations of Eq.~\eqref{eq:gr-angles} are not native to hardware, and their decomposition into elementary gates dominates the cost. The M\"ott\"onen scheme~\cite{Mottonen_2004} removes them entirely. Each step $j$ is a uniformly controlled rotation, that is, a multiplexer that applies $R_y(\theta_l^{(j)})$ conditioned on the value of the $j-1$ control qubits. It is decomposed into $2^{j-1}$ single-qubit $R_y$ rotations along with $2^{j-1}$ CNOT gates, whose controls follow the Gray code, in which consecutive integers differ in a single bit. The angles are replaced by
\begin{equation}
  \tilde\theta_i=2^{-(j-1)}\sum_{l}(-1)^{\,b_l\cdot g_i}\,\theta_l ,
\end{equation}
where $b_l$ and $g_i$ are the binary and Gray-code representations of the indices, a Walsh--Hadamard-type transform of the Grover--Rudolph angles. Summed over the $m$ steps, the circuit has at most $2^m-1$ rotations and $2^m-2$ CNOTs, i.e.\ $\Theta(M)$ gates, and no multi-controlled gates. The decomposition is exact. Since in HoLCU the Hadamard qubit also controls $\mathrm{PREP}$ and $\mathrm{PREP}^\dagger$, it is included as an additional control for each multiplexer, and the same decomposition applies, yielding roughly twice as many rotations and CNOTs.

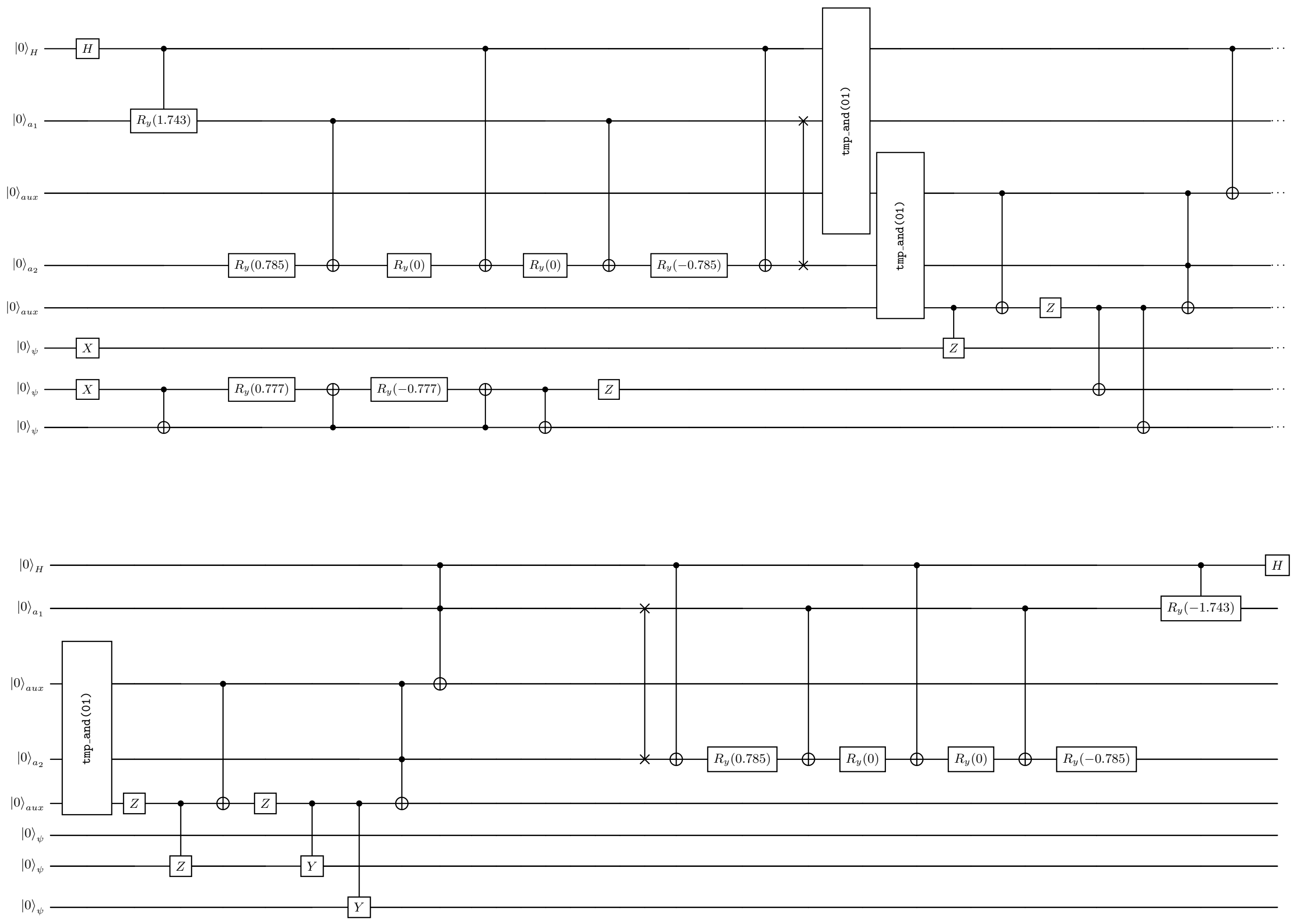
\begin{figure*}[t]
\centering

\resizebox{0.99\textwidth}{!}{%
\begin{quantikz}[row sep=0.5cm, column sep=0.8cm]
\lstick{$\ket{0}_H$}
& \gate{H} & \ctrl{1} & \qw & \qw & \qw & \ctrl{3} & \qw & \qw & \qw & \ctrl{3} & \qw
& \gate[3,style={minimum width=1.2cm}]{\rotatebox{90}{\texttt{tmp\_and(01)}}}
& \qw & \qw & \qw & \qw & \qw & \qw & \qw & \ctrl{2}
& \push{\cdots} \\
\lstick{$\ket{0}_{a_1}$}
& \qw & \gate{R_y(1.743)} & \qw & \ctrl{2} & \qw & \qw & \qw & \ctrl{2} & \qw & \qw & \swap{2}
& \ctrl{1} & \qw & \qw & \qw & \qw & \qw & \qw & \qw & \qw
& \push{\cdots} \\
\lstick{$\ket{0}_{aux}$}
& \qw & \qw & \qw & \qw & \qw & \qw & \qw & \qw & \qw & \qw & \qw
& \qw
& \gate[3,style={minimum width=1.2cm}]{\rotatebox{90}{\texttt{tmp\_and(01)}}}
& \qw & \ctrl{2} & \qw & \qw & \qw & \ctrl{1} & \targ{}
& \push{\cdots} \\
\lstick{$\ket{0}_{a_2}$}
& \qw & \qw & \gate{R_y(0.785)} & \targ{} & \gate{R_y(0)} & \targ{} & \gate{R_y(0)} & \targ{} & \gate{R_y(-0.785)} & \targ{} & \targX{}
& \qw & \ghost{\texttt{tmp\_and(01)}} & \qw & \qw & \qw & \qw & \qw & \ctrl{1} & \qw
& \push{\cdots} \\
\lstick{$\ket{0}_{aux}$}
& \qw & \qw & \qw & \qw & \qw & \qw & \qw & \qw & \qw & \qw & \qw
& \qw & \ghost{\texttt{tmp\_and}} & \ctrl{1} & \targ{} & \gate{Z} & \ctrl{2} & \ctrl{3} & \targ{} & \qw
& \push{\cdots} \\
\lstick{$\ket{0}_{\psi}$}
& \gate{X} & \qw & \qw & \qw & \qw & \qw & \qw & \qw & \qw & \qw & \qw
& \qw & \qw & \gate{Z} & \qw & \qw & \qw & \qw & \qw & \qw
& \push{\cdots} \\
\lstick{$\ket{0}_{\psi}$}
& \gate{X} & \ctrl{1} & \gate{R_y(0.777)} & \targ{} & \gate{R_y(-0.777)} & \targ{} & \ctrl{1} & \gate{Z} & \qw & \qw & \qw
& \qw & \qw & \qw & \qw & \qw & \targ{} & \qw & \qw & \qw
& \push{\cdots} \\
\lstick{$\ket{0}_{\psi}$}
& \qw & \targ{} & \qw & \ctrl{-1} & \qw & \ctrl{-1} & \targ{} & \qw & \qw & \qw & \qw
& \qw & \qw & \qw & \qw & \qw & \qw & \targ{} & \qw & \qw
& \push{\cdots}
\end{quantikz}%
}
\vspace{1.5cm}

\resizebox{0.99\textwidth}{!}{%
\begin{quantikz}[row sep=0.5cm, column sep=0.6cm]
\lstick{$\ket{0}_H$}
& \qw & \qw & \qw & \qw & \qw & \qw & \qw & \qw & \ctrl{1} & \qw & \qw
& \qw & \qw & \qw & \qw & \qw
& \qw & \ctrl{3} & \qw & \qw & \qw & \ctrl{3} & \qw & \qw & \qw & \ctrl{1} & \gate{H} \\
\lstick{$\ket{0}_{a_1}$}
& \qw & \qw & \qw & \qw & \qw & \qw & \qw & \qw & \ctrl{1} & \qw & \qw
& \qw & \qw & \qw & \qw & \qw
& \swap{2} & \qw & \qw & \ctrl{2} & \qw & \qw & \qw & \ctrl{2} & \qw & \gate{R_y(-1.743)} & \qw \\
\lstick{$\ket{0}_{aux}$}
& \gate[3,style={minimum width=1.2cm}]{\rotatebox{90}{\texttt{tmp\_and(01)}}}
& \qw & \qw & \ctrl{2} & \qw & \qw & \qw & \ctrl{1} & \targ{} & \qw & \qw
& \qw & \qw & \qw & \qw & \qw
& \qw & \qw & \qw & \qw & \qw & \qw & \qw & \qw & \qw & \qw & \qw \\
\lstick{$\ket{0}_{a_2}$}
& \ghost{\texttt{tmp\_and}}
& \qw & \qw & \qw & \qw & \qw & \qw & \ctrl{1} & \qw & \qw & \qw
& \qw & \qw & \qw & \qw & \qw
& \targX{} & \targ{} & \gate{R_y(0.785)} & \targ{} & \gate{R_y(0)} & \targ{} & \gate{R_y(0)} & \targ{} & \gate{R_y(-0.785)} & \qw & \qw \\
\lstick{$\ket{0}_{aux}$}
& \ghost{\texttt{tmp\_and}}
& \gate{Z} & \ctrl{2} & \targ{} & \gate{Z} & \ctrl{2} & \ctrl{3} & \targ{} & \qw & \qw & \qw
& \qw & \qw & \qw & \qw & \qw
& \qw & \qw & \qw & \qw & \qw & \qw & \qw & \qw & \qw & \qw & \qw \\
\lstick{$\ket{0}_{\psi}$}
& \qw & \qw & \qw & \qw & \qw & \qw & \qw & \qw & \qw & \qw & \qw
& \qw & \qw & \qw & \qw & \qw
& \qw & \qw & \qw & \qw & \qw & \qw & \qw & \qw & \qw & \qw & \qw \\
\lstick{$\ket{0}_{\psi}$}
& \qw & \qw & \gate{Z} & \qw & \qw & \gate{Y} & \qw  & \qw & \qw & \qw & \qw
& \qw & \qw & \qw & \qw & \qw
& \qw & \qw & \qw & \qw & \qw & \qw & \qw & \qw & \qw & \qw & \qw \\
\lstick{$\ket{0}_{\psi}$}
& \qw & \qw & \qw & \qw & \qw & \qw & \gate{Y} & \qw & \qw & \qw & \qw
& \qw & \qw & \qw & \qw & \qw
& \qw & \qw & \qw & \qw & \qw & \qw & \qw & \qw & \qw & \qw & \qw
\end{quantikz}%
}

\caption{Decomposed HoLCU circuit for the calculation of the $\mathcal{M}_{E2}$ reduced matrix element of $^6$He, consisting only of 1-, and 2-qubit gates, and Toffoli gates. The $\mathrm{PREP}$ part is implemented using the M\"ott\"onen state preparation method~\cite{Mottonen_2004}, and the multi-controlled gates of the $\mathrm{SELECT}$ part are decomposed using the unary iteration method~\cite{Babbush_2018}. The unary iteration builds them from single-controlled gates, Toffoli gates, and temporary logical-AND gates (Fig.~\ref{fig:tmp_and}), which require two auxiliary qubits, $\ket{0}_{\mathrm{aux}}$. These qubits are returned to $\ket{0}$ at the end of the $\mathrm{SELECT}$ part. Negative coefficients are incorporated by applying a $Z$-gate on the last auxiliary qubit.}
\label{fig:decomposed_6He}
\end{figure*}

\subsection{Unary iteration method for SELECT operator} \label{SELECT}
The $\mathrm{SELECT}$ operator of Eq.~\eqref{eq:select} applies $\xi_kU_k$ to the state register when the LCU register is in $\ket{k}$. The raw implementation is a sequence of $M$ operations, each controlled by $m$ qubits of the index register, which requires every multi-controlled gate to recompute the same Boolean logic even though it changes only slowly from one index to the next, for a cost of $O(Mm)$ Toffoli-type gates.
Unary iteration, introduced by~\cite{Babbush_2018}, avoids this repetition by going through the indices $k=0,1,2,\dots$ in order and reusing the constructs from the previous step. Moving from $k$ to $k+1$ changes only the partial results that depend on the bits that differ between the two indices; only these are updated with a few gates, rather than recomputing the full control. At the end of each chain, a flag qubit equals $1$ if and only if the register holds $\ket{k}$, and it controls a single-controlled $\xi_kU_k$. The partial results are computed on auxiliary qubits known to start in $\ket{0}$, so the AND gates that compute them can be replaced by cheaper temporary-AND (``elbow'') gates~\cite{Gidney_2018}. For $M$ target operators and $m=\lceil\log_2M\rceil$ control qubits, the method uses $m-1$ clean auxiliary qubits. This method notably reduces circuit depth through multi-controlled operations, at the expense of introducing auxiliary qubits.

\begin{figure*}[t]
    \centering

    \begin{subfigure}[t]{0.49\textwidth}
        \includegraphics[width=\textwidth]{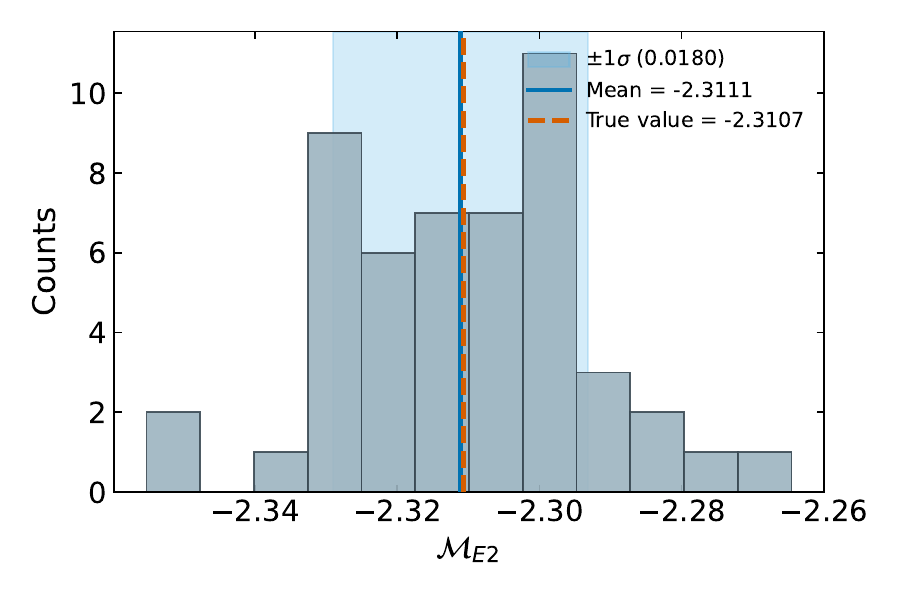}
        \label{fig:sub1}
    \end{subfigure}
    \begin{subfigure}[t]{0.49\textwidth}
        \includegraphics[width=\textwidth]{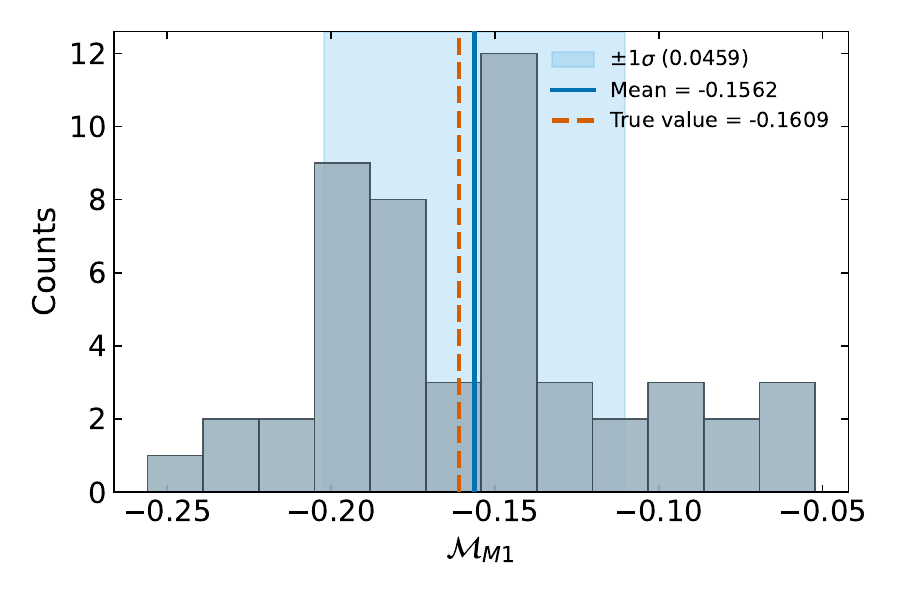}
        \label{fig:sub2}
    \end{subfigure}
    \caption{\small Histogram of the reduced matrix elements $\mathcal{M}_{E2}$ and $\mathcal{M}_{M1}$ for $^6$He, obtained from noiseless simulations with 50 independent runs of 10000 shots each. The solid blue line marks the mean, and the shaded band shows the standard deviation around it. The dashed orange line marks the exact value from classical calculations.}

    \label{fig:6He_E2_M1}
\end{figure*}

\begin{table*}[t]
\centering
\renewcommand{\arraystretch}{1.35}

\caption{Quantum circuit resources for the HoLCU circuit used to calculate the
$\mathcal{M}_{E2}$ and $\mathcal{M}_{M1}$ reduced matrix elements for the $2^+_1$ state of $^6$He, comparing the raw and reduced circuit implementations.}
\label{tab:circuit_resources_He}

\begin{ruledtabular}
\begin{tabular}{c|cc|cc|cc|cc|cc}
\textbf{Observable}
& \multicolumn{2}{c|}{\textbf{Qubits}}
& \multicolumn{2}{c|}{\textbf{Gate count}}
& \multicolumn{2}{c|}{\textbf{2Q gates}}
& \multicolumn{2}{c|}{\textbf{1Q gates}} 
& \multicolumn{2}{c}{\textbf{Circuit depth}} \\

& \textbf{Raw} & \textbf{Reduced}
& \textbf{Raw} & \textbf{Reduced}
& \textbf{Raw} & \textbf{Reduced}
& \textbf{Raw} & \textbf{Reduced}
& \textbf{Raw} & \textbf{Reduced} \\

\hline

$\mathcal{M}_{E2}$
& 6 & 8
& 962 & 114
& 410 & 46
& 552 & 68 
& 755 & 79 \\

$\mathcal{M}_{M1}$
& 7 & 10
& 2466 & 237
& 1062 & 102
& 1404 & 135
& 1953 & 176 \\

\end{tabular}
\end{ruledtabular}
\end{table*}

\begin{table*}[t]
\centering
\renewcommand{\arraystretch}{1.3}

\caption{Comparison of the exact and error-mitigated reduced matrix elements $\mathcal{M}_{E2}$ and $\mathcal{M}_{M1}$ for $^{6}$He, obtained from noisy simulations and quantum hardware calculations. Results are presented for linear and exponential zero-noise extrapolation (ZNE), both with and without readout error mitigation (REM).}
\label{tab:final_results_6He}

\begin{ruledtabular}
\begin{tabular}{cccccc}

\textbf{Observable}
& \textbf{Exact}
& \multicolumn{2}{c}{\textbf{ZNE}}
& \multicolumn{2}{c}{\textbf{ZNE + REM}}
\\

&
& \textbf{Linear}
& \textbf{Exponential}
& \textbf{Linear}
& \textbf{Exponential}
\\

\hline

$\mathcal{M}_{E2}$ (Noisy sim.)
& $-2.3106$
& $-2.0996 \pm 0.0092$
& $-2.1665 \pm 0.0111$
& $-2.2093 \pm 0.0097$
& $-2.279 \pm 0.0116$

\\

$\mathcal{M}_{E2}$ (Hardware)
& $-2.3106$
& $-1.5239 \pm 0.0077$
& $-$
& $-1.5681 \pm 0.0078$
& $-$

\\
\hline

$\mathcal{M}_{M1}$ (Noisy sim.)
& $-0.1609$
& $-0.3269 \pm 0.0017$
& $-0.2475 \pm 0.0024$
& $-0.3063 \pm 0.0018$
& $-0.2201 \pm 0.0025$
\\

$\mathcal{M}_{M1}$ (Hardware)
& $-0.1609$
& $-0.986 \pm 0.0137$
& $-$
& $-0.984 \pm 0.0137$
& $-$
\\

\end{tabular}
\end{ruledtabular}
\end{table*}

\begin{table*}[t]
\centering
\renewcommand{\arraystretch}{1.35}

\caption{Quantum circuit resources for the HoLCU circuit used to calculate the $\mathcal{M}_{E2}$ and $\mathcal{M}_{M1}$ reduced matrix elements for the $2^+_1$ state of $^{18}$O, comparing the raw and reduced circuit implementations.}
\label{tab:circuit_resources_18O}

\begin{ruledtabular}
\begin{tabular}{c|cc|cc|cc|cc|cc}
\textbf{Observable}
& \multicolumn{2}{c|}{\textbf{Qubits}}
& \multicolumn{2}{c|}{\textbf{Gate count}}
& \multicolumn{2}{c|}{\textbf{2Q gates}}
& \multicolumn{2}{c|}{\textbf{1Q gates}}
& \multicolumn{2}{c}{\textbf{Circuit depth}} \\

& \textbf{Raw} & \textbf{Reduced}
& \textbf{Raw} & \textbf{Reduced}
& \textbf{Raw} & \textbf{Reduced}
& \textbf{Raw} & \textbf{Reduced}
& \textbf{Raw} & \textbf{Reduced} \\

\hline

$\mathcal{M}_{E2}$
& 15 & 20
& 139378 & 1310
& 59681 & 608
& 79697 & 702
& 111381 & 841 \\

$\mathcal{M}_{M1}$
& 14 & 18
& 28817 & 779
& 12152 & 346
& 16665 & 433
& 22345 & 501 \\

\end{tabular}
\end{ruledtabular}
\end{table*}

\section{RESULTS AND DISCUSSION} \label{results}

In this section, we present the results for the reduced matrix elements ($\mathcal{M}_{M1}$ and $\mathcal{M}_{E2}$) corresponding to the magnetic dipole and electric quadrupole moments of $^6$He and $^{18}$O, calculated using the HoLCU and HoLCU-div techniques. We report results from noiseless and noisy simulations for both nuclei, and hardware results only for the $^6$He observables. We also discuss the decomposition of the raw circuits, which contain many multi-controlled gates in the $\mathrm{PREP}$ and $\mathrm{SELECT}$ sections. The reduced circuits consist only of 1- and 2-qubit gates; this comes at the expense of additional qubits, but it decreases the circuit depth by orders of magnitude. The wave function parameters are obtained from noiseless simulations and then used in the noisy simulations and the real quantum hardware calculations. The noisy simulations are performed using the Qiskit~\cite{JavadiAbhari_2024} \texttt{FakeFez} backend, and the actual quantum computer calculations use the \textit{ibm\_fez} hardware. For quantum error mitigation in the noisy simulations and hardware calculations, we employ zero-noise extrapolation and readout-error mitigation.

\begin{figure*}[!htbp]
    \centering

    \begin{subfigure}[t]{0.49\textwidth}
        \includegraphics[width=\textwidth]{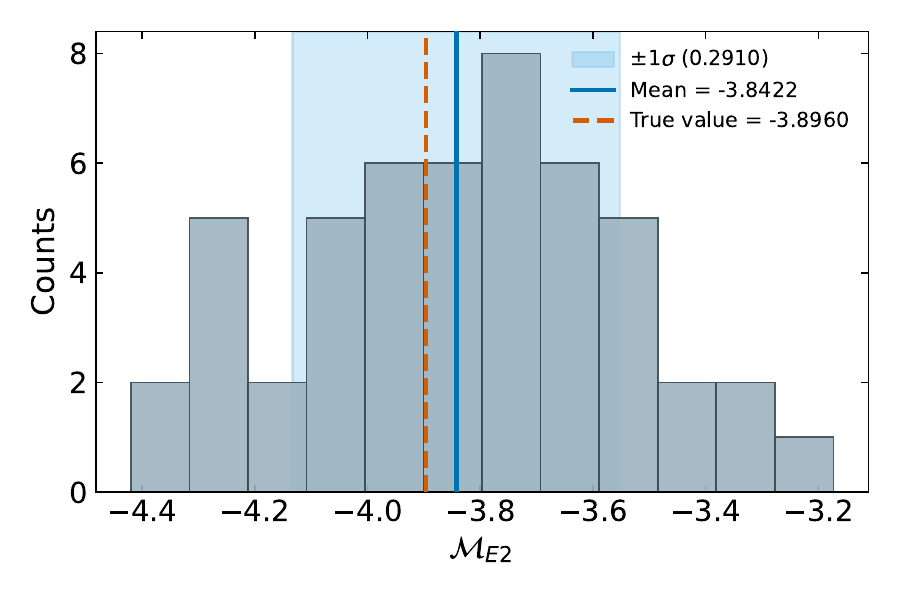}
        \label{fig:sub1}
    \end{subfigure}
    \begin{subfigure}[t]{0.49\textwidth}
        \includegraphics[width=\textwidth]{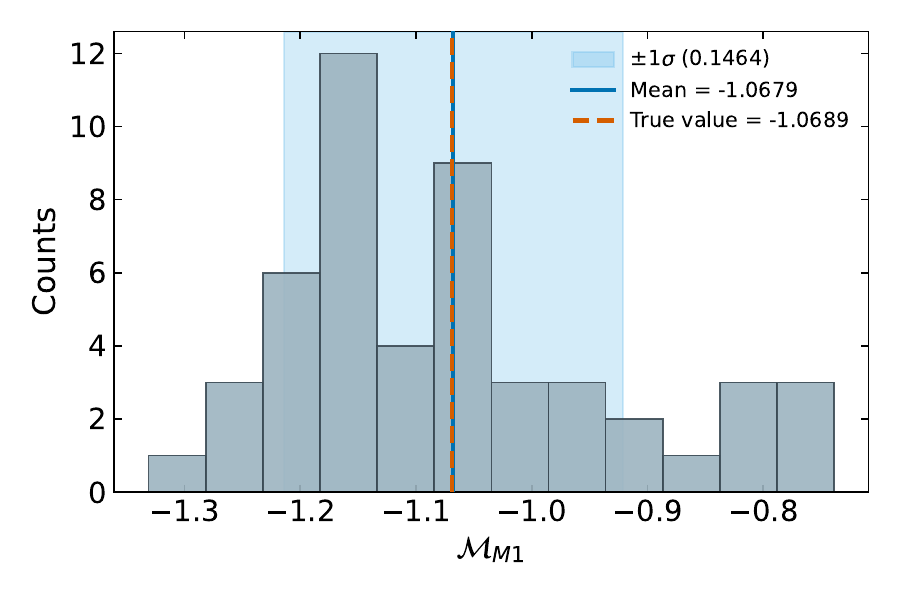}
        \label{fig:sub2}
    \end{subfigure}
    \caption{\small Histogram of the reduced matrix elements $\mathcal{M}_{E2}$ and $\mathcal{M}_{M1}$ for $^{18}$O, obtained from noiseless simulations with 50 independent runs of 10000 shots each. The solid blue line marks the mean, and the shaded band shows the standard deviation around it. The dashed orange line marks the exact value from classical calculations.}

    \label{fig:18O_E2_M1}
\end{figure*}

\subsection{$^6$He}
The first excited state of $^6$He (2$^+$) is represented as a 3-qubit VQE circuit as discussed in the previous section. On the other hand, the $\mathcal{M}_{E2}$ operator can be expanded into four Pauli strings, which can be represented using a two-qubit LCU ancilla register. Including the Hadamard qubit, the circuit for evaluating $\mathcal{M}_{E2}$ becomes a 6-qubit circuit before any decomposition is applied. The raw quantum circuit for computing $\mathcal{M}_{E2}$ is shown in Fig.~\ref{fig:Raw_6He}; it represents the HoLCU technique used in our calculations. The blue dashed box marks the $\mathrm{PREP}$ part of the circuit, and the orange dashed box marks the $\mathrm{SELECT}$ part. The red region is the wave function, with parameters obtained from noiseless VQE simulations. The operators are written in Pauli string format (Table~\ref{tab:pauli_coefficients_6He_E2} in Appendix~\ref{app:appendixa}), using the JW encoding. Negative coefficients are incorporated in the raw circuit by applying an appropriate combination of gates on the ancilla qubits of the LCU section within the $\mathrm{SELECT}$ part. Similarly, a 7-qubit circuit can be constructed for evaluating $\mathcal{M}_{M1}$ for the same state. Now, the $\mathcal{M}_{M1}$ operator can be represented by six Pauli strings as shown in Table~\ref{tab:pauli_coefficients_6He_M1} from Appendix~\ref{app:appendixa}, which can be represented by 3 qubits.

The final decomposed circuits for both $\mathcal{M}_{E2}$ operator is shown in Fig.~\ref{fig:decomposed_6He}. It decomposes the multi-controlled gates at the expense of introducing auxiliary qubits: two for $\mathcal{M}_{E2}$ and three for $\mathcal{M}_{M1}$. These auxiliary qubits are a consequence of using the unary iteration method. In this circuit, negative coefficients are incorporated by applying a $Z$-gate on the last auxiliary qubit just before the corresponding Pauli string is applied. 
In Table~\ref{tab:circuit_resources_He}, we show a comparison of resource counts required for evaluating $\mathcal{M}_{E2}$ and $\mathcal{M}_{M1}$ following the raw circuits and the final decomposed circuits. The table shows a clear reduction in the number of 2-qubit gates and depth, which makes them more suitable for current quantum hardware. 
For $\mathcal{M}_{E2}$, the 2-qubit gate count drops from 410 to 46, and
for $\mathcal{M}_{M1}$ from 1062 to 102, an order of magnitude lower in both cases. On the other hand, the depth of the decomposed circuits is also significantly reduced compared to the raw circuits.

To verify our method, we first ran our raw and decomposed circuits within a noiseless simulator. Fig.~\ref {fig:6He_E2_M1} shows the noiseless simulation for reduced matrix elements of $\mathcal{M}_{E2}$ and $\mathcal{M}_{M1}$. These results are obtained from 50 independent runs with 10000 shots each, and their close agreement with the exact results gives us confidence that the constructed circuits are the correct ones.

Finally, the decomposed circuits for both observables were executed using a noisy simulator (FakeFez) and a quantum hardware ($ibm\_fez$).
The results are shown in Table~\ref{tab:final_results_6He}, where the noisy simulations used 50 independent runs with 10000 shots each and hardware calculations used 10 independent runs with 10000 shots each for both observables.
We show results for linear and exponential extrapolations in the zero-noise extrapolation (ZNE), both as standalone mitigation techniques and when combined with readout-error mitigation (REM). In the noisy simulations, ZNE along with REM and an exponential fit yields the best performance for both observables. The hardware results in Table~\ref{tab:final_results_6He} are far from the exact values for both observables. For $\mathcal{M}_{E2}$, the linear ZNE result deviates from the exact value by about $34\%$. Adding readout-error mitigation reduces the deviation to about $32\%$. For $\mathcal{M}_{M1}$, the deviation is much
\begin{table*}[t]
\centering
\renewcommand{\arraystretch}{1.35}

\caption{Quantum circuit resources for the HoLCU-div circuits to  calculate the
$\mathcal{M}_{E2}$ reduced matrix elements for the $2^+_1$ state of $^{18}$O.}
\label{tab:HoLCU-div_resources_18OE2}

\begin{ruledtabular}
\begin{tabular}{cccccc}
\textbf{Coefficient}
& \textbf{Qubits}
& \textbf{Gate count}
& \textbf{2Q gates}
& \textbf{1Q gates}
& \textbf{Circuit depth} \\

\hline

$1.4153$ & 13 & 300 & 127 & 173 & 156 \\
$1.6175$ & 11 & 270 & 114 & 156 & 153 \\
$0.4043$ & 10 & 247 & 105 & 142 & 155 \\
$2.0219$ & 10 & 248 & 105 & 143 & 157 \\
$0.4952$ & 13 & 365 & 160 & 205 & 207 \\
$1.2131$ & 11 & 310 & 134 & 176 & 171 \\
$1.8084$ & 13 & 433 & 196 & 237 & 238 \\
$2.2149$ & 13 & 381 & 169 & 212 & 217 \\

\end{tabular}
\end{ruledtabular}
\end{table*}
\begin{table*}[t]
\centering
\renewcommand{\arraystretch}{1.35}

\caption{Quantum circuit resources for the HoLCU-div circuits to  calculate the
$\mathcal{M}_{M1}$ reduced matrix elements for the $2^+_1$ state of $^{18}$O.}
\label{tab:HoLCU-div_resources_18OM1}

\begin{ruledtabular}
\begin{tabular}{cccccc}
\textbf{Coefficient}
& \textbf{Qubits}
& \textbf{Gate count}
& \textbf{2Q gates}
& \textbf{1Q gates}
& \textbf{Circuit depth} \\

\hline

$0.2559$ & 13 & 315 & 133 & 182 & 168 \\
$0.7679$ & 11 & 267 & 113 & 154 & 168 \\
$1.2799$ & 13 & 301 & 127 & 174 & 176 \\
$1.2541$ & 13 & 363 & 160 & 203 & 205 \\
$1.0239$ & 11 & 310 & 134 & 176 & 171 \\

\end{tabular}
\end{ruledtabular}
\end{table*}
larger: the linear ZNE and ZNE + REM values are more than five times the exact magnitude. Readout-error mitigation makes almost no difference in this case. In both cases, the statistical uncertainties are much smaller than the deviations from the exact values, which indicates that the discrepancy is dominated by hardware noise rather than shot noise. The larger deviation for $\mathcal{M}_{M1}$ is consistent with its deeper circuit (102 2-qubit gates compared to 46 for $\mathcal{M}_{E2}$), and with the fact that $\mathcal{M}_{M1}$ is much smaller in magnitude, so a similar absolute error gives a much larger relative error.
The exponential extrapolations are not shown for the hardware results. When the noise is amplified by gate-folding, the expectation value changes sign at a noise scale factor of 5; because the exponential form cannot capture this sign change, the fit is not possible.

\subsection{$^{18}$O}
Similar to the case of $^6$He, the first excited state of $^{18}$O which is again a 2$^+$ state, can be defined using a 9-qubit circuit. On the other hand, the $\mathcal{M}_{E2}$ and $\mathcal{M}_{M1}$ operators can be written as 22 and 16 Pauli strings, respectively, as shown in Tables~\ref{tab:pauli_coefficients_18O_E2} and \ref{tab:pauli_coefficients_18O_M1}. The PREP block can be constructed using 5 qubits for $\mathcal{M}_{E2}$ and 4 qubits for $\mathcal{M}_{M1}$. Overall, a 15-qubit circuit is needed for evaluating the $\mathcal{M}_{E2}$ reduced matrix element and a 14-qubit circuit for the $\mathcal{M}_{M1}$ reduced matrix element, including one Hadamard qubit for both cases.
Table~\ref{tab:circuit_resources_18O} shows the resources of the raw and decomposed forms of the full HoLCU circuit. The decomposition of the multi-controlled gates comes at the expense of introducing auxiliary qubits: five for $\mathcal{M}_{E2}$ and four for $\mathcal{M}_{M1}$. In return, the circuit depth is reduced from 111381 to 841 for $\mathcal{M}_{E2}$ and from 22345 to 501 for $\mathcal{M}_{M1}$, which are reductions by factors of about 132 and 45, respectively. The number of 2-qubit gates, which are the main source of noise on quantum hardware, is reduced from 59681 to 608 for $\mathcal{M}_{E2}$, and from 12152 to 346 for $\mathcal{M}_{M1}$, again by factors of 98 and 35, respectively.

We first performed a noiseless simulation using the reduced circuits for both observables. Fig.~\ref{fig:18O_E2_M1} shows the noiseless simulation results performed using 50 independent runs with 10000 shots each for both reduced matrix elements, using the full HoLCU circuit. Good agreement with the exact values gives us confidence that the constructed circuits work perfectly. The noisy simulations are performed using the HoLCU-div method with 50 independent runs, each with 10000 shots\footnote{The expectation values corresponding to the coefficients $0.4952$ and $2.0219$ in Table~\ref{tab:groups_E2_18O} were evaluated using $10^6$ and $10^5$ shots, respectively, due to their small magnitudes. For all other coefficients, $10^4$ shots were used.}, in which Pauli strings are grouped by the magnitude of their coefficients, and each group is evaluated with its own
\begin{table*}[t]
\centering
\renewcommand{\arraystretch}{1.3}

\caption{Comparison of the error-mitigated expectation values of the Pauli strings associated with each coefficient in the HoLCU-div implementation for calculating the $\mathcal{M}_{E2}$ reduced matrix element of $^{18}$O. Results obtained using linear and exponential ZNE are presented, both with and without readout error mitigation (REM).}
\label{tab:groups_E2_18O}

\begin{ruledtabular}
\begin{tabular}{cccccc}

\textbf{Coefficients}
& \textbf{Operators}
& \multicolumn{2}{c}{\textbf{ZNE}}
& \multicolumn{2}{c}{\textbf{ZNE + REM}}
\\

& 
& \textbf{Linear}
& \textbf{Exponential}
& \textbf{Linear}
& \textbf{Exponential}
\\

\hline

\multirow{3}{*}{$1.4153$}
& $-Z_0$ &  &  &  &  \\
& $-Z_1$ & $-1.0295 \pm 0.0146$ & $-1.0332 \pm 0.0155$ & $-1.0387 \pm 0.0154$ & $-1.0427 \pm 0.0164$ \\
& $Z_2$  &  &  &  &  \\

\hline

\multirow{2}{*}{$1.6175$}
& $-Z_3$ & $-0.7992 \pm 0.0077$ & $-0.8017 \pm 0.0083$ & $-0.8419 \pm 0.0082$ & $-0.8445 \pm 0.0087$ \\
& $-Z_4$ &  &  &  &  \\

\hline

$0.4043$
& $-Z_5$ & $-0.1174 \pm 0.0037$ & $-0.1174 \pm 0.0037$ & $-0.1263 \pm 0.0039$ & $-0.1263 \pm 0.0038 $ \\

\hline

$2.0219$
& $Z_6$ & $-0.0113 \pm 0.0003$ & $-0.0113 \pm 0.0003$ & $-0.0144 \pm 0.0004$ & $-0.0144 \pm 0.0004$ \\

\hline

\multirow{4}{*}{$0.4952$}
& $Y_0 Z_1 Z_2 Y_3$ &  &  &  &  \\
& $X_0 Z_1 Z_2 X_3$ & $0.0001 \pm 0.0014$ & $0.0017 \pm 0.4711$ & $-0.0089 \pm 0.0014$ & $-0.0117 \pm 0.0165$ \\
& $-Y_1 Z_2 Z_3 Y_4$ &  &  &  &  \\
& $-X_1 Z_2 Z_3 X_4$ &  &  &  &  \\

\hline

\multirow{2}{*}{$1.2131$}
& $-Y_2 Z_3 Z_4 Y_5$ & $-0.1628 \pm 0.0071$ & $-0.2072 \pm 0.0137$ & $-0.1721 \pm 0.0074$ & $-0.2171 \pm 0.0140$ \\
& $-X_2 Z_3 Z_4 X_5$ &  &  &  &  \\

\hline

\multirow{4}{*}{$1.8084$}
& $-Y_0 Z_1 \cdots Z_6 Y_7$ &  &  &  &  \\
& $-X_0 Z_1 \cdots Z_6 X_7$ & $-0.0859 \pm 0.0150$ & $-0.1034 \pm 0.0316$ & $-0.1001 \pm 0.0158$ & $-0.1150 \pm 0.0268$ \\
& $Y_1 Z_2 \cdots Z_7 Y_8$ &  &  &  &  \\
& $X_1 Z_2 \cdots Z_7 X_8$ &  &  &  &  \\

\hline

\multirow{4}{*}{$2.2149$}
& $-Y_3 Z_4 Z_5 Z_6 Y_7$ &  &  &  &  \\
& $-X_3 Z_4 Z_5 Z_6 X_7$ & $-1.1644 \pm 0.0124$ & $-1.3921 \pm 0.0208$ & $-1.2278 \pm 0.0131$ & $-1.4674 \pm 0.0219$ \\
& $-Y_4 Z_5 Z_6 Z_7 Y_8$ &  &  &  &  \\
& $-X_4 Z_5 Z_6 Z_7 X_8$ &  &  &  &  \\

\end{tabular}
\end{ruledtabular}
\end{table*}
circuit. In HoLCU-div, the $\mathcal{M}_{M1}$ operator has 16 Pauli strings, which form five groups, and the $\mathcal{M}_{E2}$ operator has 22 Pauli strings, which form eight groups. The largest group contains four strings, so every circuit needs at most two LCU ancilla qubits and one auxiliary qubit, and its $\mathrm{PREP}$ part consists only of controlled-Hadamard gates. The resources of all the circuits are listed in Table~\ref{tab:HoLCU-div_resources_18OE2} for $\mathcal{M}_{E2}$ and in Table~\ref{tab:HoLCU-div_resources_18OM1} for $\mathcal{M}_{M1}$. The largest circuit for $\mathcal{M}_{E2}$ uses 13 qubits, 196 2-qubit gates, and has a depth of 238, compared with 20 qubits, 608 2-qubit gates, and a depth of 841 for the decomposed full circuit. Similarly, the largest circuit for $\mathcal{M}_{M1}$ uses 13 qubits, 160 2-qubit gates, and has a depth of 205, compared with 18 qubits, 346 2-qubit gates, and a depth of 501. The price is that $M_t$ number of circuits have to be run instead of one. The largest circuits for both reduced matrix elements are presented in Appendix~\ref{app:appendixc}.

The group-wise error-mitigated values are given in Table~\ref{tab:groups_E2_18O} for $\mathcal{M}_{E2}$ and in Table~\ref{tab:groups_M1_18O} for $\mathcal{M}_{M1}$.The final value of each observable is obtained by summing the group contributions, $|c_g|\,\mathrm{Re}\langle B_g\rangle$, and adding the constant (identity) term of the operator, which is given in Tables~\ref{tab:pauli_coefficients_18O_E2} and~\ref{tab:pauli_coefficients_18O_M1} of Appendix~\ref{app:appendixb}. For the group with coefficient $0.4952$ in $\mathcal{M}_{E2}$, the signal is consistent with zero and the exponential fit is poorly constrained, which can be seen from its large uncertainty.

Finally, in Table~\ref{tab:final_results_18O}, we present the noisy simulation results for both observables. We have shown results for linear and exponential extrapolations for zero-noise extrapolation (ZNE) and for ZNE combined with readout-error mitigation (REM). For $\mathcal{M}_{E2}$, the linear ZNE value of $-2.7281 \pm 0.0464$ deviates from the exact value of $-3.8960$ by about $28\%$, which improves to about $23\%$ with REM. The exponential fit gives $-3.3264 \pm 0.2465$, which gives a deviation of about $15\%$, and ZNE along with REM and an exponential fit gives $-3.6254 \pm 0.0761$, which is within about $7\%$ of the exact value. For $\mathcal{M}_{M1}$, the linear ZNE value of $-1.2686 \pm 0.0259$ deviates from the exact value of $-1.0689$ by about $19\%$ ($14\%$ with REM), while ZNE along with REM and an exponential fit gives $-1.1700 \pm 0.0370$, a deviation of about $9\%$. The results show that ZNE, along with readout-error mitigation and an exponential fit,
\begin{table*}[t]
\centering
\renewcommand{\arraystretch}{1.3}

\caption{Comparison of the error-mitigated expectation values of the Pauli strings associated with each coefficient in the HoLCU-div implementation for calculating the $\mathcal{M}_{M1}$ reduced matrix element of $^{18}$O. Results obtained using linear and exponential ZNE are presented, both with and without readout error mitigation (REM).}
\label{tab:groups_M1_18O}

\begin{ruledtabular}
\begin{tabular}{cccccc}

\textbf{Coefficients}
& \textbf{Operators}
& \multicolumn{2}{c}{\textbf{ZNE}}
& \multicolumn{2}{c}{\textbf{ZNE + REM}}
\\

& 
& \textbf{Linear}
& \textbf{Exponential}
& \textbf{Linear}
& \textbf{Exponential}
\\

\hline
\multirow{4}{*}{$0.2559$}
& $Z_0$ &  &  &  &  \\
& $-Z_1$ & $-0.3010 \pm 0.0143$ & $-0.3383 \pm 0.0216$ & $-0.3050 \pm 0.0150$ & $-0.3471 \pm 0.0233$ \\
& $-Z_3$ &  &  &  &  \\
& $Z_4$ &   &  &  &  \\

\hline

\multirow{2}{*}{$0.7679$}
& $-Z_2$ & $-0.6692 \pm 0.0056$ & $-0.7246 \pm 0.0077$ & $-0.7052 \pm 0.0059$ & $-0.7626 \pm 0.0080$ \\
& $Z_5$ &   &  &  &  \\

\hline

\multirow{3}{*}{$1.2799$}
& $Z_6$ &  &  &  &  \\
& $-Z_7$ & $0.2238 \pm 0.0131$ & $0.2565 \pm 0.0153$ & $0.2701 \pm 0.0138$ & $0.3053 \pm 0.0162$ \\
& $Z_8$ &  &  &  &  \\

\hline

\multirow{4}{*}{$1.2541$}
& $Y_0Z_1Z_2Y_3$ &  &  &  &  \\
& $X_0Z_1Z_2X_3$ & $0.1182 \pm 0.0139$ & $0.1284 \pm 0.0195$ & $0.1249 \pm 0.0145$ & $0.1354 \pm 0.0204$ \\
& $Y_1Z_2Z_3Y_4$ &  &  &  &  \\
& $X_1Z_2Z_3X_4$ &   &  &  &  \\

\hline

\multirow{2}{*}{$1.0239$}
& $X_2Z_3Z_4X_5$ & $0.1636 \pm 0.0070$ & $0.2080 \pm 0.0135$ & $ 0.1721 \pm 0.0074$ & $0.2185 \pm 0.0141$ \\
& $Y_2Z_3Z_4Y_5$ &  &  &  &  \\

\end{tabular}
\end{ruledtabular}
\end{table*}
\begin{table*}[t]
\centering
\renewcommand{\arraystretch}{1.3}

\caption{Comparison of the exact and error-mitigated reduced matrix elements $\mathcal{M}_{E2}$ and $\mathcal{M}_{M1}$ for $^{18}$O, obtained from noisy simulations using the HoLCU-div circuits. Results are presented for linear and exponential zero-noise extrapolation (ZNE), both with and without readout error mitigation (REM).}
\label{tab:final_results_18O}

\begin{ruledtabular}
\begin{tabular}{cccccc}

\textbf{Observable}
& \textbf{Exact}
& \multicolumn{2}{c}{\textbf{ZNE}}
& \multicolumn{2}{c}{\textbf{ZNE + REM}}
\\

&
& \textbf{Linear}
& \textbf{Exponential}
& \textbf{Linear}
& \textbf{Exponential}
\\

\hline

$\mathcal{M}_{E2}$ (HoLCU-div)
& $-3.8960$
& $-2.7281 \pm 0.0464$
& $-3.3264 \pm 0.2465$
& $-3.0019 \pm 0.0489$
& $-3.6254 \pm 0.0761$

\\

$\mathcal{M}_{M1}$ (HoLCU-div)
& $-1.0689$
& $-1.2686 \pm 0.0259$
& $-1.2206 \pm 0.0352$
& $-1.2209 \pm 0.0271$
& $-1.1700 \pm 0.0370$

\\

\end{tabular}
\end{ruledtabular}
\end{table*}
yields the best performance for both observables.


\section{Summary and Conclusion} \label{summary}

We have calculated the reduced matrix elements $\mathcal{M}_{M1}$ and $\mathcal{M}_{E2}$ of $^6$He and $^{18}$O using the HoLCU and HoLCU-div techniques. The wave function parameters were obtained from noiseless VQE simulations and then used in the noisy simulations and, for $^6$He, in hardware calculations on \textit{ibm\_fez}. Zero-noise extrapolation and readout-error mitigation techniques were used for all noisy and hardware calculations.

The raw HoLCU circuits contain many multi-controlled gates in the $\mathrm{PREP}$ and $\mathrm{SELECT}$ sections, and we have decomposed them into circuits consisting only of 1- and 2-qubit gates. This comes at the expense of additional auxiliary qubits arising from the unary iteration method, but it reduces circuit complexity by orders of magnitude. In contrast to the standard LCU approach, HoLCU yields the expectation value of the full operator directly from measurements of the Hadamard-encoded qubit, without any post-selection on the ancilla register.

For $^{18}$O, the HoLCU-div variant reduces the circuit depth further when the coefficients are degenerate: the largest circuits use 13 qubits, 196 and 160 2-qubit gates, and depths of 238 and 205 for $\mathcal{M}_{E2}$ and $\mathcal{M}_{M1}$, respectively, compared with 20 and 18 qubits, 608 and 346 2-qubit gates, and depths of 841 and 501 for the decomposed full circuits. This comes at the cost of running multiple circuits.

In the noisy simulations, ZNE, together with readout-error mitigation and an exponential fit, yields the best results for both observables and nuclei. For $^{18}$O, the best results deviate from the exact values by about $7\%$ for $\mathcal{M}_{E2}$ and about $9\%$ for $\mathcal{M}_{M1}$. The hardware results for $^6$He deviate significantly from the exact values, especially for $\mathcal{M}_{M1}$, indicating that hardware noise remains a limiting factor even for the reduced circuits. Exponential extrapolation could not be applied there because the values change sign when amplified by noise.

Although in the present work we focus mainly on two-nucleon systems, the employed HoLCU techniques offer many potential extensions for implementing shell-model calculations using quantum algorithms. Firstly, using compact encodings to represent nuclear wave functions and expanding the observables in the same encoding can reduce the number of qubits needed, making the circuits more reliable on current-generation quantum hardware. At the same time, it has the potential to scale to other complex nuclear systems beyond simple two-nucleon systems.  Secondly, this method can be extended to dynamic observables as well, such as electromagnetic transitions between two different states of a single nucleus or even for $\beta$-decay processes, which represent transitions between states of two different nuclei. \\

\begin{acknowledgments}
We acknowledge financial support from MHRD (India) and Research Grant No. ANRF/ARGM/2025/001130/TS from Anusandhan National Research Foundation [ANRF], India.
C.S. acknowledges UK Science and Technology Facilities Council (STFC) grant ST/Y000358/1.
R.M.S. would like to thank the organizers and lecturers of the ECT*-TALENT school ``Quantum Computing for Nuclear Physics'', held at the European Center for Theoretical Studies in Nuclear Physics and Related Areas (ECT*), whose lectures and discussions helped him to shape and inspire the work presented in this paper. He is especially grateful to M.~Hjorth-Jensen and A.~Roggero for fruitful discussions throughout this summer school.
\end{acknowledgments}





\vspace{0.5cm}

\onecolumngrid

\appendix

\section{Transformed nuclear observables for $^{6}\mathrm{He}$ } \label{app:appendixa}

This appendix presents the Pauli-string decompositions of the transformed nuclear observables $\mathcal{M}_{E2}$ and $\mathcal{M}_{M1}$ for $^6$He, obtained by mapping the corresponding operators to qubit operators using the JW encoding, as given in Tables~\ref{tab:pauli_coefficients_6He_E2}--\ref{tab:pauli_coefficients_6He_M1}. These representations serve as the input operators for their subsequent implementation in the HoLCU quantum circuits.

\begin{table}[h]
\centering
\renewcommand{\arraystretch}{1.3}

\caption{Pauli-string decomposition of the $\mathcal{M}_{E2}$ operator for $^6$He, together with the corresponding expansion coefficients.}
\label{tab:pauli_coefficients_6He_E2}

\begin{ruledtabular}
\begin{tabular}{cc cc}
\textbf{Operator} & \textbf{Coefficient}
& \textbf{Operator} & \textbf{Coefficient} \\
\hline
$Z_0$    & $0.60625537$
& $Z_1$    & $-0.60625537$ \\

$X_1X_2$ & $-0.85737456$
& $Y_1Y_2$ & $-0.85737456$
\end{tabular}
\end{ruledtabular}
\end{table}

\begin{table}[h]
\centering
\renewcommand{\arraystretch}{1.3}

\caption{Pauli-string decomposition of the $\mathcal{M}_{M1}$ operator for $^6$He, together with the corresponding expansion coefficients.}
\label{tab:pauli_coefficients_6He_M1}

\begin{ruledtabular}
\begin{tabular}{cccc}
\textbf{Operator} & \textbf{Coefficient}
& \textbf{Operator} & \textbf{Coefficient} \\
\hline
$I$      & $-1.27996045$
& $Z_0$    & $1.27995166$ \\

$Z_1$    & $0.42665055$
& $X_1X_2$ & $1.20674999$ \\

$Y_1Y_2$ & $1.20674999$
& $Z_2$    & $-0.42664177$
\end{tabular}
\end{ruledtabular}
\end{table}

\FloatBarrier

\newpage
\section{Transformed nuclear observables for $^{18}$O} \label{app:appendixb}

This appendix presents the Pauli-string decompositions of the transformed nuclear observables $\mathcal{M}_{E2}$ and $\mathcal{M}_{M1}$ for $^{18}$O, obtained by mapping the corresponding operators to qubit operators using the JW encoding. The decompositions, along with their associated expansion coefficients, are provided in Tables~\ref{tab:pauli_coefficients_18O_E2}--\ref{tab:pauli_coefficients_18O_M1}. These representations serve as the input operators for their subsequent implementation in the HoLCU quantum circuits.

\begin{table}[h]
\centering
\renewcommand{\arraystretch}{1.3}

\caption{Pauli-string decomposition of the $\mathcal{M}_{E2}$ operator for $^{18}$O, together with the corresponding expansion coefficients.}
\label{tab:pauli_coefficients_18O_E2}

\begin{ruledtabular}
\begin{tabular}{cccc}
\textbf{Operator} & \textbf{Coefficient}
& \textbf{Operator} & \textbf{Coefficient} \\
\hline
$I$ & $3.03284923$
& $Z_0$ & $-1.41532868$ \\

$Z_1$ & $-1.41532868$
& $Z_2$ & $1.41532868$ \\

$Z_3$ & $-1.61752055$
& $Z_4$ & $-1.61752055$ \\

$Z_5$ & $-0.40438014$
& $Z_6$ & $2.02190069$ \\

$Y_0Z_1Z_2Y_3$
& $0.49525000$
& $X_0Z_1Z_2X_3$
& $0.49525000$ \\

$Y_1Z_2Z_3Y_4$
& $-0.49525000$
& $X_1Z_2Z_3X_4$
& $-0.49525000$ \\

$Y_2Z_3Z_4Y_5$
& $-1.21310980$
& $X_2Z_3Z_4X_5$
& $-1.21310980$ \\

$Y_0Z_1Z_2Z_3Z_4Z_5Z_6Y_7$
& $-1.80843716$
& $X_0Z_1Z_2Z_3Z_4Z_5Z_6X_7$
& $-1.80843716$ \\

$Y_1Z_2Z_3Z_4Z_5Z_6Z_7Y_8$
& $1.80843716$
& $X_1Z_2Z_3Z_4Z_5Z_6Z_7X_8$
& $1.80843716$ \\

$Y_3Z_4Z_5Z_6Y_7$
& $-2.21489071$
& $X_3Z_4Z_5Z_6X_7$
& $-2.21489071$ \\

$Y_4Z_5Z_6Z_7Y_8$
& $-2.21489071$
& $X_4Z_5Z_6Z_7X_8$
& $-2.21489071$
\end{tabular}
\end{ruledtabular}
\end{table}

\begin{table}[h]
\centering
\renewcommand{\arraystretch}{1.3}

\caption{Pauli-string decomposition of the $\mathcal{M}_{M1}$ operator for $^{18}$O, together with the corresponding expansion coefficients.}
\label{tab:pauli_coefficients_18O_M1}

\begin{ruledtabular}
\begin{tabular}{cccc}
\textbf{Operator} & \textbf{Coefficient}
& \textbf{Operator} & \textbf{Coefficient} \\
\hline
$I$      & $-1.27999170$
& $Z_0$    & $0.25599033$ \\

$Z_1$    & $-0.25599033$
& $Z_2$    & $-0.76797100$ \\

$Z_3$    & $-0.25599534$
& $Z_4$    & $0.25599534$ \\

$Z_5$    & $0.76798601$
& $Z_6$    & $1.27997669$ \\

$Z_7$    & $-1.27998121$
& $Z_8$    & $1.27998121$ \\

$Y_0Z_1Z_2Y_3$
& $1.25409139$
& $X_0Z_1Z_2X_3$
& $1.25409139$ \\

$Y_1Z_2Z_3Y_4$
& $1.25409139$
& $X_1Z_2Z_3X_4$
& $1.25409139$ \\

$Y_2Z_3Z_4Y_5$
& $1.02396133$
& $X_2Z_3Z_4X_5$
& $1.02396133$
\end{tabular}
\end{ruledtabular}
\end{table}

 \clearpage

\section{Circuit building blocks and largest HoLCU-div circuits for $^{18}$O}
\label{app:appendixc}  

This appendix contains the quantum circuit diagrams used in the HoLCU and HoLCU-div implementations of the $\mathcal{M}_{E2}$ and $\mathcal{M}_{M1}$ matrix elements. It shows the elementary building blocks, such as the temporary logical-AND gate used in the unary iteration and the uniform initialization of the LCU ancilla qubits used in the HoLCU-div circuits, together with the largest HoLCU-div circuits for calculating both observables of $^{18}$O.

\begin{figure}[h]
  \centering
  \resizebox{\columnwidth}{!}{%
  \begin{quantikz}[row sep=0.5cm, column sep=0.8cm]
  \lstick{$\ket{1}$} & \gate[3, style={minimum width=1.2cm}]{\rotatebox{90}{\texttt{tmp\_and(01)}}} & \qw \\
  \lstick{$\ket{0}$} & \ghost{\texttt{tmp\_and}} & \qw \\
  \lstick{$\ket{0}$} & \ghost{\texttt{tmp\_and}} & \qw
  \end{quantikz}
  =\begin{quantikz}[row sep=0.5cm, column sep=0.4cm]
  \lstick{$\ket{1}$} & \qw & \qw & \qw & \qw & \ctrl{2} & \qw & \qw & \qw & \qw & \qw \\
  \lstick{$\ket{0}$} & \qw & \qw & \ctrl{1} & \qw & \qw & \qw & \ctrl{1} & \qw & \qw & \qw \\
  \lstick{$\ket{0}$} & \gate{H} & \gate{T} & \targ{} & \gate{T^\dagger} & \targ{} & \gate{T} & \targ{} & \gate{T^\dagger} & \gate{H} & \gate{S^\dagger}
  \end{quantikz}%
  }
  \caption{Decomposition of the temporary logical-AND gate into Clifford and $T$ gates. The gate computes the logical AND of the two control qubits into a third qubit that starts in $\ket{0}$; the third qubit is always the auxiliary qubit. It is used in the unary iteration method to build the multi-controlled gates of the $\mathrm{SELECT}$ part from single-controlled gates~\cite{Babbush_2018}.}
  \label{fig:tmp_and}
\end{figure}
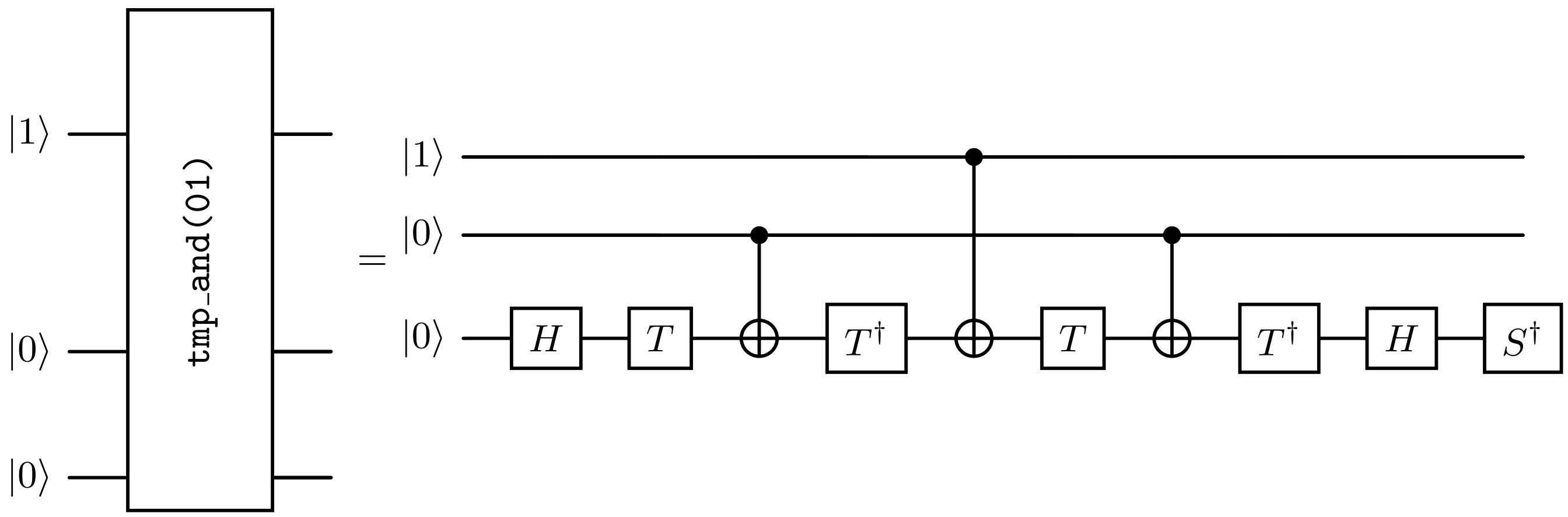

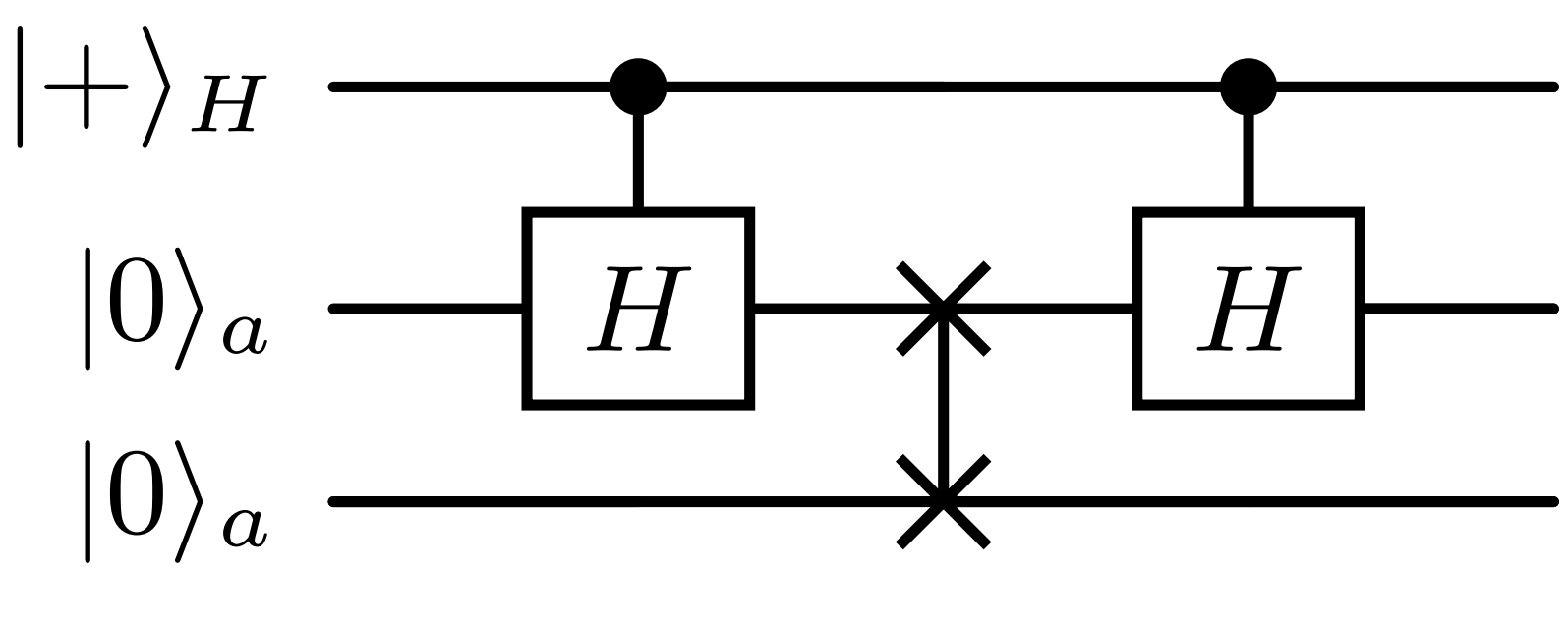
\begin{figure}[h]
\centering
\resizebox{0.6\textwidth}{!}{%
\begin{quantikz}[row sep=0.25cm, column sep=0.5cm]
\lstick{$|+\rangle_H$}
& \ctrl{1} & \qw & \ctrl{1} & \qw
\\
\lstick{$|0\rangle_a$}
& \gate{H} & \swap{1} & \gate{H} & \qw
\\
\lstick{$|0\rangle_a$}
& \qw & \targX{} & \qw & \qw
\\
\end{quantikz}%
}
\caption{Quantum circuit for uniform initialization with two ancilla qubits used in the HoLCU-div circuit for the $\mathrm{PREP}$ part of the circuit.}
\end{figure}

\begin{figure*}[ht]
\centering
\resizebox{\textwidth}{!}{%
\begin{quantikz}[row sep=0.50cm, column sep=0.35cm]
\lstick{$\ket{0}_H$}
& \gate{H} & \ctrl{1} & \qw & \ctrl{1} & \qw & \qw & \ctrl{5} & \ctrl{6} & \ctrl{7} & \ctrl{8} & \ctrl{9} & \ctrl{10} & \ctrl{11} & \ctrl{12} & \qw & \qw & \qw & \qw & \qw & \qw & \qw & \qw & \qw & \qw & \qw & \qw & \qw & \qw & \qw & \qw & \qw & \qw & \qw & \qw & \qw & \qw & \qw & \qw & \qw & \qw & \qw & \qw & \qw & \qw & \qw & \ctrl{1} & \qw & \ctrl{1} & \gate{H} & \meter{} \\
\lstick{$\ket{0}_{a_1}$}
& \qw & \gate{H} & \swap{1} & \gate{H} & \qw & \octrl{2} & \qw & \qw & \qw & \qw & \qw & \qw & \qw & \qw & \octrl{2} & \qw & \qw & \qw & \qw & \qw & \qw & \qw & \qw & \qw & \ctrl{2} & \qw & \qw & \qw & \qw & \qw & \qw & \qw & \qw & \qw & \ctrl{2} & \qw & \qw & \qw & \qw & \qw & \qw & \qw & \qw & \qw & \ctrl{2} & \gate{H} & \swap{1} & \gate{H} & \qw \\
\lstick{$\ket{0}_{a_2}$}
& \qw & \qw & \targX{} & \qw & \qw & \octrl{1} & \qw & \qw & \qw & \qw & \qw & \qw & \qw & \qw & \qw & \qw & \qw & \qw & \qw & \qw & \qw & \qw & \qw & \ctrl{1} & \qw & \qw & \qw & \qw & \qw & \qw & \qw & \qw & \qw & \qw & \qw & \qw & \qw & \qw & \qw & \qw & \qw & \qw & \qw & \qw & \ctrl{1} & \qw & \targX{} & \qw & \qw \\
\lstick{$\ket{0}_{aux}$}
& \qw & \qw & \qw & \qw & \qw & \targ{} & \ctrl{2} & \ctrl{3} & \ctrl{4} & \ctrl{5} & \ctrl{6} & \ctrl{7} & \ctrl{8} & \ctrl{9} & \targ{} & \ctrl{2} & \ctrl{3} & \ctrl{4} & \ctrl{5} & \ctrl{6} & \ctrl{7} & \ctrl{8} & \ctrl{9} & \targ{} & \targ{} & \gate{Z} & \ctrl{1} & \ctrl{2} & \ctrl{3} & \ctrl{4} & \ctrl{5} & \ctrl{6} & \ctrl{7} & \ctrl{8} & \targ{} & \gate{Z} & \ctrl{1} & \ctrl{2} & \ctrl{3} & \ctrl{4} & \ctrl{5} & \ctrl{6} & \ctrl{7} & \ctrl{8} & \targ{} & \qw & \qw & \qw & \qw \\
\lstick{$\ket{0}_{\psi}$}
& \qw & \qw & \qw & \qw & \gate[9,style={minimum width=1.5cm}]{\rotatebox{90}{\texttt{O18\_wav}}} & \qw & \qw & \qw & \qw & \qw & \qw & \qw & \qw & \qw & \qw & \qw & \qw & \qw & \qw & \qw & \qw & \qw & \qw & \qw & \qw & \qw & \gate{Y} & \qw & \qw & \qw & \qw & \qw & \qw & \qw & \qw & \qw & \targ{} & \qw & \qw & \qw & \qw & \qw & \qw & \qw & \qw & \qw & \qw & \qw & \qw  \\
\lstick{$\ket{0}_{\psi}$}
& \qw & \qw & \qw & \qw & \ghost{\texttt{O18\_wav}} & \qw & \gate{Y} & \qw & \qw & \qw & \qw & \qw & \qw & \qw & \qw & \targ{} & \qw & \qw & \qw & \qw & \qw & \qw & \qw & \qw & \qw & \qw & \qw & \gate{Z} & \qw & \qw & \qw & \qw & \qw & \qw & \qw & \qw & \qw & \gate{Z} & \qw & \qw & \qw & \qw & \qw & \qw & \qw & \qw & \qw & \qw & \qw \\
\lstick{$\ket{0}_{\psi}$}
& \qw & \qw & \qw & \qw & \ghost{\texttt{O18\_wav}} & \qw & \qw & \gate{Z} & \qw & \qw & \qw & \qw & \qw & \qw & \qw & \qw & \gate{Z} & \qw & \qw & \qw & \qw & \qw & \qw & \qw & \qw & \qw & \qw & \qw & \gate{Z} & \qw & \qw & \qw & \qw & \qw & \qw & \qw & \qw & \qw & \gate{Z} & \qw & \qw & \qw & \qw & \qw & \qw & \qw & \qw & \qw & \qw \\
\lstick{$\ket{0}_{\psi}$}
& \qw & \qw & \qw & \qw & \ghost{\texttt{O18\_wav}} & \qw & \qw & \qw & \gate{Z} & \qw & \qw & \qw & \qw & \qw & \qw & \qw & \qw & \gate{Z} & \qw & \qw & \qw & \qw & \qw & \qw & \qw & \qw & \qw & \qw & \qw & \gate{Z} & \qw & \qw & \qw & \qw & \qw & \qw & \qw & \qw & \qw & \gate{Z} & \qw & \qw & \qw & \qw & \qw & \qw & \qw & \qw & \qw \\
\lstick{$\ket{0}_{\psi}$}
& \qw & \qw & \qw & \qw & \ghost{\texttt{O18\_wav}} & \qw & \qw & \qw & \qw & \gate{Z} & \qw & \qw & \qw & \qw & \qw & \qw & \qw & \qw & \gate{Z} & \qw & \qw & \qw & \qw & \qw & \qw & \qw & \qw & \qw & \qw & \qw & \gate{Z} & \qw & \qw & \qw & \qw & \qw & \qw & \qw & \qw & \qw & \gate{Z} & \qw & \qw & \qw & \qw & \qw & \qw & \qw & \qw \\
\lstick{$\ket{0}_{\psi}$}
& \qw & \qw & \qw & \qw & \ghost{\texttt{O18\_wav}} & \qw & \qw & \qw & \qw & \qw & \gate{Z} & \qw & \qw & \qw & \qw & \qw & \qw & \qw & \qw & \gate{Z} & \qw & \qw & \qw & \qw & \qw & \qw & \qw & \qw & \qw & \qw & \qw & \gate{Z} & \qw & \qw & \qw & \qw & \qw & \qw & \qw & \qw & \qw & \gate{Z} & \qw & \qw & \qw & \qw & \qw & \qw & \qw \\
\lstick{$\ket{0}_{\psi}$}
& \qw & \qw & \qw & \qw & \ghost{\texttt{O18\_wav}} & \qw & \qw & \qw & \qw & \qw & \qw & \gate{Z} & \qw & \qw & \qw & \qw & \qw & \qw & \qw & \qw & \gate{Z} & \qw & \qw & \qw & \qw & \qw & \qw & \qw & \qw & \qw & \qw & \qw & \gate{Z} & \qw & \qw & \qw & \qw & \qw & \qw & \qw & \qw & \qw & \gate{Z} & \qw & \qw & \qw & \qw & \qw & \qw  \\
\lstick{$\ket{0}_{\psi}$}
& \qw & \qw & \qw & \qw & \ghost{\texttt{O18\_wav}} & \qw & \qw & \qw & \qw & \qw & \qw & \qw & \gate{Z} & \qw & \qw & \qw & \qw & \qw & \qw & \qw & \qw & \gate{Z} & \qw & \qw & \qw & \qw & \qw & \qw & \qw & \qw & \qw & \qw & \qw & \gate{Y} & \qw & \qw & \qw & \qw & \qw & \qw & \qw & \qw & \qw & \targ{} & \qw & \qw & \qw & \qw & \qw \\
\lstick{$\ket{0}_{\psi}$}
& \qw & \qw & \qw & \qw & \ghost{\texttt{O18\_wav}} & \qw & \qw & \qw & \qw & \qw & \qw & \qw & \qw & \gate{Y} & \qw & \qw & \qw & \qw & \qw & \qw & \qw & \qw & \targ{} & \qw & \qw & \qw & \qw & \qw & \qw & \qw & \qw & \qw & \qw & \qw & \qw & \qw & \qw & \qw & \qw & \qw & \qw & \qw & \qw & \qw & \qw & \qw & \qw & \qw & \qw 
\end{quantikz}%
}
\caption{Largest HoLCU-div circuit for the calculation of $\mathcal{M}_{E2}$ of $^{18}$O. It corresponds to the group with coefficient $|c_g|=1.8084$, which contains four Pauli strings with the same magnitude of the coefficient, and therefore uses two LCU ancilla qubits. The multi-controlled gates of the $\mathrm{SELECT}$ part are decomposed with the unary iteration method, and the signs of the terms are applied by a $Z$-gate on the last auxiliary qubit.}
\end{figure*}
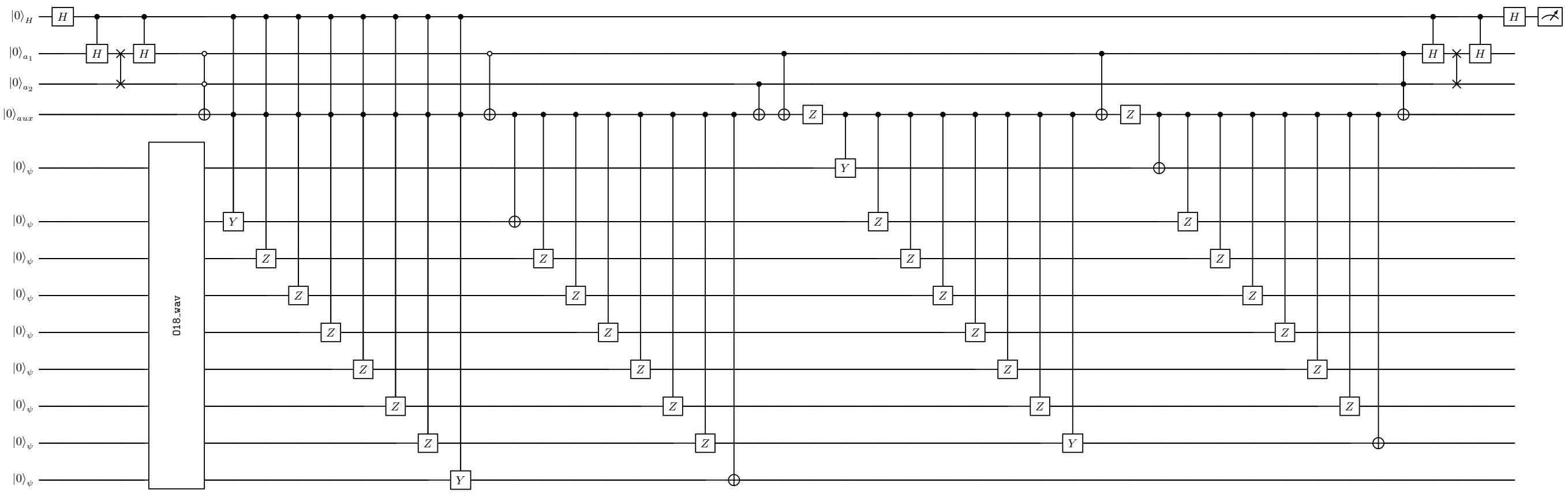

\begin{minipage}{\textwidth}
  \centering
  \resizebox{\textwidth}{!}{%
    \begin{quantikz}[row sep=0.35cm, column sep=0.35cm]
        \lstick{$\ket{0}_H$}
        & \gate{H} & \ctrl{1} & \qw & \ctrl{1} & \qw & \qw & \ctrl{4} & \ctrl{5} & \ctrl{6} & \ctrl{7} & \qw & \qw & \qw & \qw & \qw & \qw & \qw & \qw & \qw & \qw & \qw & \qw & \qw & \qw & \qw & \qw & \qw & \ctrl{1} & \qw & \ctrl{1} & \gate{H} & \meter{} \\
        \lstick{$\ket{0}_{a_1}$}
        & \qw & \gate{H} & \swap{1} & \gate{H} & \qw & \octrl{2} & \qw & \qw & \qw & \qw & \octrl{2} & \qw & \qw & \qw & \qw & \qw & \ctrl{2} & \qw & \qw & \qw & \qw & \ctrl{2} & \qw & \qw & \qw & \qw & \ctrl{1} & \gate{H} & \swap{1} & \gate{H} & \qw \\
        \lstick{$\ket{0}_{a_2}$}
        & \qw & \qw & \targX{} & \qw & \qw & \octrl{1} & \qw & \qw & \qw & \qw & \qw & \qw & \qw & \qw & \ctrl{1} & \ctrl{1} & \qw & \qw & \qw & \qw & \qw & \qw & \qw & \qw & \qw & \ctrl{1} & \ctrl{1} & \qw & \targX{} & \qw & \qw \\
        \lstick{$\ket{0}_{aux}$}
        & \qw & \qw & \qw & \qw & \qw & \targ{} & \ctrl{1} & \ctrl{2} & \ctrl{3} & \ctrl{4} & \targ{} & \ctrl{1} & \ctrl{2} & \ctrl{3} & \ctrl{4} & \targ{} & \targ{} & \ctrl{2} & \ctrl{3} & \ctrl{4} & \ctrl{5} & \targ{} & \ctrl{2} & \ctrl{3} & \ctrl{4} & \ctrl{5} & \targ{} & \qw & \qw & \qw & \qw \\
        \lstick{$\ket{0}_{\psi}$}
        & \qw & \qw & \qw & \qw & \gate[9,style={minimum width=1.5cm}]{\rotatebox{90}{\texttt{O18\_wav}}} & \qw & \gate{Y} & \qw & \qw & \qw & \qw & \targ{} & \qw & \qw & \qw & \qw & \qw & \qw & \qw & \qw & \qw & \qw & \qw & \qw & \qw & \qw & \qw & \qw & \qw & \qw & \qw \\
        \lstick{$\ket{0}_{\psi}$}
        & \qw & \qw & \qw & \qw & \ghost{\texttt{O18\_wav}} & \qw & \qw & \gate{Z} & \qw & \qw & \qw & \qw & \gate{Z} & \qw & \qw & \qw & \qw & \gate{Y} & \qw & \qw & \qw & \qw & \targ{} & \qw & \qw & \qw & \qw & \qw & \qw & \qw & \qw \\
        \lstick{$\ket{0}_{\psi}$}
        & \qw & \qw & \qw & \qw & \ghost{\texttt{O18\_wav}} & \qw & \qw & \qw & \gate{Z} & \qw & \qw & \qw & \qw & \gate{Z} & \qw & \qw & \qw & \qw & \gate{Z} & \qw & \qw & \qw & \qw & \gate{Z} & \qw & \qw & \qw & \qw & \qw & \qw & \qw \\
        \lstick{$\ket{0}_{\psi}$}
        & \qw & \qw & \qw & \qw & \ghost{\texttt{O18\_wav}} & \qw & \qw & \qw & \qw & \gate{Y} & \qw & \qw & \qw & \qw & \targ{} & \qw & \qw & \qw & \qw & \gate{Z} & \qw & \qw & \qw & \qw & \gate{Z} & \qw & \qw & \qw & \qw & \qw & \qw \\
        \lstick{$\ket{0}_{\psi}$}
        & \qw & \qw & \qw & \qw & \ghost{\texttt{O18\_wav}} & \qw & \qw & \qw & \qw & \qw & \qw & \qw & \qw & \qw & \qw & \qw & \qw & \qw & \qw & \qw & \gate{Y} & \qw & \qw & \qw & \qw & \targ{}  & \qw & \qw & \qw & \qw & \qw \\
        \lstick{$\ket{0}_{\psi}$}
        & \qw & \qw & \qw & \qw & \ghost{\texttt{O18\_wav}} & \qw & \qw & \qw & \qw & \qw & \qw & \qw & \qw & \qw & \qw & \qw & \qw & \qw & \qw & \qw & \qw & \qw & \qw & \qw & \qw & \qw & \qw & \qw & \qw & \qw & \qw \\
        \lstick{$\ket{0}_{\psi}$}
        & \qw & \qw & \qw & \qw & \ghost{\texttt{O18\_wav}} & \qw & \qw & \qw & \qw & \qw & \qw & \qw & \qw & \qw & \qw & \qw & \qw & \qw & \qw & \qw & \qw & \qw & \qw & \qw & \qw & \qw & \qw & \qw & \qw & \qw & \qw \\
        \lstick{$\ket{0}_{\psi}$}
        & \qw & \qw & \qw & \qw & \ghost{\texttt{O18\_wav}} & \qw & \qw & \qw & \qw & \qw & \qw & \qw & \qw & \qw & \qw & \qw & \qw & \qw & \qw & \qw & \qw & \qw & \qw & \qw & \qw & \qw & \qw & \qw & \qw & \qw & \qw
    \end{quantikz}%
  }
  \captionof{figure}{Largest HoLCU-div circuit for the calculation of $\mathcal{M}_{M1}$ of $^{18}$O. It corresponds to the group with coefficient $|c_g|=1.2541$, which contains four Pauli strings with the same magnitude of the coefficient, and therefore uses two LCU ancilla qubits. The multi-controlled gates of the $\mathrm{SELECT}$ part are decomposed with the unary iteration method, and the signs of the terms are applied by a $Z$-gate on the last auxiliary qubit.}
  \label{fig:holcu_div_M1_O18}
\end{minipage}

\bigskip    

\newpage
\twocolumngrid   
\bibliography{references}

@book{chuang,
  author    = {Nielsen, M. A. and Chuang, I. L.},
  title     = {Quantum Computation and Quantum Information},
  edition   = {10},
  publisher = {Cambridge University Press},
  year      = {2010}
}

@article{nisq,
  author  = {Preskill, J.},
  title   = {Quantum Computing in the {NISQ} era and beyond},
  journal = {Quantum},
  volume  = {2},
  pages   = {79},
  year    = {2018},
  doi     = {10.22331/q-2018-08-06-79}
}

@article{Klco_2022,
  author  = {Klco, N. and Roggero, A. and Savage, M. J.},
  title   = {Standard Model Physics and the Digital Quantum Revolution: Thoughts about the Interface},
  journal = {Rep. Prog. Phys.},
  volume  = {85},
  number  = {6},
  pages   = {064301},
  year    = {2022},
  doi     = {10.1088/1361-6633/ac58a4}
}

@article{vqe,
  author  = {{A. Peruzzo, J. McClean, P. Shadbolt et al.}},
  title   = {A variational eigenvalue solver on a photonic quantum processor},
  journal = {Nat. Commun.},
  volume  = {5},
  pages   = {4213},
  year    = {2014},
  doi     = {10.1038/ncomms5213}
}

@article{Perez2_2023,
  author  = {P{\'e}rez-Obiol, A. and Romero, A. M. and Men{\'e}ndez, J. and Rios, A. and Garc{\'i}a-S{\'a}ez, A. and Juli{\'a}-D{\'i}az, B.},
  title   = {Nuclear shell-model simulation in digital quantum computers},
  journal = {Sci. Rep.},
  volume  = {13},
  pages   = {12291},
  year    = {2023},
  doi     = {10.1038/s41598-023-39263-7}
}

@article{Romero_2022,
  author  = {Romero, A. M. and Engel, J. and Tang, H. L. and Economou, S. E.},
  title   = {Solving Nuclear Structure Problems with the Adaptive Variational Quantum Algorithm},
  journal = {Phys. Rev. C},
  volume  = {105},
  number  = {6},
  pages   = {064317},
  year    = {2022},
  doi     = {10.1103/PhysRevC.105.064317}
}

@article{Lacroix_PRL,
  author  = {Lacroix, D.},
  title   = {Symmetry-assisted preparation of entangled many-body states on a quantum computer},
  journal = {Phys. Rev. Lett.},
  volume  = {125},
  pages   = {230502},
  year    = {2020},
  doi     = {10.1103/PhysRevLett.125.230502}
}

@article{Lee,
  title = {Rodeo Algorithm for Quantum Computing},
  author = {Choi, Kenneth and Lee, Dean and Bonitati, Joey and Qian, Zhengrong and Watkins, Jacob},
  journal = {Phys. Rev. Lett.},
  volume = {127},
  issue = {4},
  pages = {040505},
  numpages = {5},
  year = {2021},
  month = {Jul},
  publisher = {American Physical Society},
  doi = {10.1103/PhysRevLett.127.040505},
  url = {https://link.aps.org/doi/10.1103/PhysRevLett.127.040505}
}

@article{James,
  title = {Multinucleon structure and dynamics via quantum computing},
  author = {Du, Weijie and Vary, James P.},
  journal = {Phys. Rev. A},
  volume = {108},
  issue = {5},
  pages = {052614},
  numpages = {24},
  year = {2023},
  month = {Nov},
  publisher = {American Physical Society},
  doi = {10.1103/PhysRevA.108.052614},
  url = {https://link.aps.org/doi/10.1103/PhysRevA.108.052614}
}

@article{James1,
title = {Systematic many-fermion Hamiltonian input scheme and spectral calculations on quantum computers},
journal = {Physics Letters B},
volume = {866},
pages = {139548},
year = {2025},
issn = {0370-2693},
doi = {https://doi.org/10.1016/j.physletb.2025.139548},
url = {https://www.sciencedirect.com/science/article/pii/S0370269325003090},
author = {Weijie Du and James P. Vary}
}

@article{Yoshida,
  title = {Accurate and precise quantum computation of valence two-neutron systems},
  author = {Yoshida, Sota and Sato, Takeshi and Ogata, Takumi and Naito, Tomoya and Kimura, Masaaki},
  journal = {Phys. Rev. C},
  volume = {109},
  issue = {6},
  pages = {064305},
  numpages = {11},
  year = {2024},
  month = {Jun},
  publisher = {American Physical Society},
  doi = {10.1103/PhysRevC.109.064305},
  url = {https://link.aps.org/doi/10.1103/PhysRevC.109.064305}
}

@article{Ayral_2023,
  author  = {Ayral, T. and Besserve, P. and Lacroix, D. and Guzman, E. A. R.},
  title   = {Quantum computing with and for many-body physics},
  journal = {Eur. Phys. J. A},
  volume  = {59},
  pages   = {227},
  year    = {2023},
  doi     = {10.1140/epja/s10050-023-01141-1}
}

@article{Chandan_2023,
  author  = {Sarma, C. and Matteo, O. D. and Abhishek, A. and Srivastava, P. C.},
  title   = {Prediction of the neutron drip line in oxygen isotopes using quantum computation},
  journal = {Phys. Rev. C},
  volume  = {108},
  pages   = {064305},
  year    = {2023},
  doi     = {10.1103/PhysRevC.108.064305}
}

@article{Robin_2023,
  author  = {Robin, Caroline E. P. and Savage, Martin J.},
  title   = {Quantum simulations in effective model spaces: Hamiltonian-learning variational quantum eigensolver using digital quantum computers and application to the Lipkin-Meshkov-Glick model},
  journal = {Phys. Rev. C},
  volume  = {108},
  number  = {2},
  pages   = {024313},
  year    = {2023},
  doi     = {10.1103/PhysRevC.108.024313}
}

@article{Bharti_Bhoy,
  author  = {Bhoy, B. and Stevenson, P.},
  title   = {Shell-model study of {$^{58}$Ni} using quantum computing algorithm},
  journal = {New J. Phys.},
  volume  = {26},
  pages   = {075001},
  year    = {2024},
  doi     = {10.1088/1367-2630/ad5756}
}

@article{QC_n1,
  author  = {Dumitrescu, E. F. and McCaskey, A. J. and Hagen, G. and Jansen, G. R. and Morris, T. D. and Papenbrock, T. and Pooser, R. C. and Dean, D. J. and Lougovski, P.},
  title   = {Cloud Quantum Computing of an Atomic Nucleus},
  journal = {Phys. Rev. Lett.},
  volume  = {120},
  pages   = {210501},
  year    = {2018},
  doi     = {10.1103/PhysRevLett.120.210501}
}

@article{Higgott_2019,
  author  = {Higgott, O. and Wang, D. and Brierley, S.},
  title   = {Variational Quantum Computation of Excited States},
  journal = {Quantum},
  volume  = {3},
  pages   = {156},
  year    = {2019},
  doi     = {10.22331/q-2019-07-01-156}
}

@article{Jiang_2023,
  author  = {Jiang, C. and Pei, J.},
  title   = {Quantum Computing of the Pairing Hamiltonian at Finite Temperature},
  journal = {Phys. Rev. C},
  volume  = {107},
  number  = {4},
  pages   = {044308},
  year    = {2023},
  doi     = {10.1103/PhysRevC.107.044308}
}

@article{Nakanishi_2019,
  author  = {Nakanishi, K. M. and Mitarai, K. and Fujii, K.},
  title   = {Subspace-Search Variational Quantum Eigensolver for Excited States},
  journal = {Phys. Rev. Res.},
  volume  = {1},
  number  = {3},
  pages   = {033062},
  year    = {2019},
  doi     = {10.1103/PhysRevResearch.1.033062}
}

@misc{Maheshwari_2026,
  author        = {Maheshwari, B. and Stevenson, P. and Van Isacker, P.},
  title         = {Full-Spectrum Quantum Simulation for the Nuclear Shell Model},
  year          = {2026},
  eprint        = {2607.08235},
  archivePrefix = {arXiv},
  primaryClass  = {nucl-th}
}

@misc{Pandey_2026,
  author        = {Pandey, D. and Singh, A. and Das, A. K. and Arumugam, P.},
  title         = {Quantum Simulation of Nuclear Shell Model Using {GCM}-Based Methods on {NISQ} Devices},
  year          = {2026},
  eprint        = {2608.01769},
  archivePrefix = {arXiv},
  primaryClass  = {nucl-th}
}

@article{Zheng_2023,
  author  = {Zheng, M. and Peng, B. and Wiebe, N. and Li, A. and Yang, X. and Kowalski, K.},
  title   = {Quantum Algorithms for Generator Coordinate Methods},
  journal = {Phys. Rev. Res.},
  volume  = {5},
  number  = {2},
  pages   = {023200},
  year    = {2023},
  doi     = {10.1103/PhysRevResearch.5.023200}
}

@article{Jordan_Wigner_1928,
  author  = {Jordan, P. and Wigner, E.},
  title   = {{\"Uber} das Paulische {\"Aquivalenzverbot}},
  journal = {Z. Phys.},
  volume  = {47},
  number  = {9--10},
  pages   = {631--651},
  year    = {1928},
  doi     = {10.1007/BF01331938}
}

@article{Siwach_2021,
  author  = {Siwach, P. and Arumugam, P.},
  title   = {Quantum Simulation of Nuclear Hamiltonian with a Generalized Transformation for Gray Code Encoding},
  journal = {Phys. Rev. C},
  volume  = {104},
  number  = {3},
  pages   = {034301},
  year    = {2021},
  doi     = {10.1103/PhysRevC.104.034301}
}

@article{Lv_2023,
  author  = {Lv, P. and Wei, S. and Xie, H.-N. and Long, G.},
  title   = {{QCSH}: A Full Quantum Computer Nuclear Shell-Model Package},
  journal = {Sci. China Phys. Mech. Astron.},
  volume  = {66},
  number  = {4},
  pages   = {240311},
  year    = {2023},
  doi     = {10.1007/s11433-022-2044-5}
}

@article{Roggero_2020,
  author  = {Roggero, A. and Gu, C. and Baroni, A. and Papenbrock, T.},
  title   = {Preparation of Excited States for Nuclear Dynamics on a Quantum Computer},
  journal = {Phys. Rev. C},
  volume  = {102},
  number  = {6},
  pages   = {064624},
  year    = {2020},
  doi     = {10.1103/PhysRevC.102.064624}
}

@article{Nigro_2025,
  author  = {Nigro, L. and Barbieri, C. and Prati, E.},
  title   = {Simulation of a Three-Nucleons System Transition on Quantum Circuits},
  journal = {Adv. Quantum Technol.},
  volume  = {8},
  number  = {5},
  pages   = {2400371},
  year    = {2025},
  doi     = {10.1002/qute.202400371}
}

@article{Sarma_2026,
  author  = {Sarma, C. and Stevenson, P. D.},
  title   = {A Low-Circuit-Depth Quantum Computing Approach to the Nuclear Shell Model},
  journal = {Discov. Quantum},
  volume  = {2},
  number  = {1},
  pages   = {6},
  year    = {2026},
  doi     = {10.1007/s44464-026-00009-9}
}

@article{Liu_2025,
  author  = {Liu, D. and Du, W. and Lin, L. and Vary, J. P. and Yang, C.},
  title   = {An Efficient Quantum Circuit for Block Encoding a Pairing Hamiltonian},
  journal = {J. Comput. Sci.},
  volume  = {85},
  pages   = {102480},
  year    = {2025},
  doi     = {10.1016/j.jocs.2024.102480}
}

@article{Costa_2026,
  author  = {Costa, E. and P{\'e}rez-Obiol, A. and Men{\'e}ndez, J. and Rios, A. and Garc{\'i}a-S{\'a}ez, A. and Juli{\'a}-D{\'i}az, B.},
  title   = {Quasiparticle Pairing Encoding of Atomic Nuclei for Quantum Annealing},
  journal = {Phys. Lett. B},
  volume  = {872},
  pages   = {140042},
  year    = {2026},
  doi     = {10.1016/j.physletb.2025.140042}
}

@article{Sarma_2026_PLB,
  title = {Qubit-efficient variational algorithm for nuclear structure},
journal = {Physics Letters B},
volume = {884},
pages = {141172},
year = {2027},
issn = {0370-2693},
doi = {https://doi.org/10.1016/j.physletb.2026.141172},
url = {https://www.sciencedirect.com/science/article/pii/S0370269326010233},
author = {Chandan Sarma and P.D. Stevenson},
}

@article{graycode,
  author  = {Di Matteo, O. and McCoy, A. and Gysbers, P. and Miyagi, T. and Woloshyn, R. M. and Navr{\'a}til, P.},
  title   = {Improving Hamiltonian encodings with the Gray code},
  journal = {Phys. Rev. A},
  volume  = {103},
  pages   = {042405},
  year    = {2021},
  doi     = {10.1103/PhysRevA.103.042405}
}

@article{Singh_2025,
  author  = {Singh, N. and Siwach, P. and Arumugam, P.},
  title   = {Advancing Quantum Simulations of the Nuclear Shell Model with Gray-Code-Based Resource-Efficient Protocols},
  journal = {Phys. Rev. C},
  volume  = {112},
  number  = {3},
  pages   = {034320},
  year    = {2025},
  doi     = {10.1103/bbkf-fjxj}
}

@article{GarciaRamos_2025,
  author  = {Garc{\'i}a-Ramos, J.-E. and S{\'a}iz, {\'A}. and Arias, J. M. and Lamata, L. and P{\'e}rez-Fern{\'a}ndez, P.},
  title   = {Nuclear Physics in the Era of Quantum Computing and Quantum Machine Learning},
  journal = {Adv. Quantum Technol.},
  volume  = {8},
  number  = {12},
  pages   = {2300219},
  year    = {2025},
  doi     = {10.1002/qute.202300219}
}

@article{pooja2,
  author  = {Siwach, P. and Arumugam, P.},
  title   = {Quantum computation of nuclear observables involving linear combinations of unitary operators},
  journal = {Phys. Rev. C},
  volume  = {105},
  pages   = {064318},
  year    = {2022},
  doi     = {10.1103/PhysRevC.105.064318}
}

@article{QC_n3,
  author  = {Stetcu, I. and Baroni, A. and Carlson, J.},
  title   = {Variational approaches to constructing the many-body nuclear ground state for quantum computing},
  journal = {Phys. Rev. C},
  volume  = {105},
  pages   = {064308},
  year    = {2022},
  doi     = {10.1103/PhysRevC.105.064308}
}

@article{QC_n4,
  author  = {Romero, A. M. and Engel, J. and Tang, H. L. and Economou, S. E.},
  title   = {Solving nuclear structure problems with the adaptive variational quantum algorithm},
  journal = {Phys. Rev. C},
  volume  = {105},
  pages   = {064317},
  year    = {2022},
  doi     = {10.1103/PhysRevC.105.064317}
}

@article{QC_n5,
  author  = {Kiss, O. and Grossi, M. and Lougovski, P. and Sanchez, F. and Vallecorsa, S. and Papenbrock, T.},
  title   = {Quantum computing of the {$^6$Li} nucleus via ordered unitary coupled clusters},
  journal = {Phys. Rev. C},
  volume  = {106},
  pages   = {034325},
  year    = {2022},
  doi     = {10.1103/PhysRevC.106.034325}
}

@article{Denis_2024,
  author  = {Zhang, J. and Lacroix, D. and Beaujeault-Taudi{\`e}re, Y.},
  title   = {Neutron-proton pairing correlations described on quantum computers},
  journal = {Phys. Rev. C},
  volume  = {110},
  pages   = {064320},
  year    = {2024},
  doi     = {10.1103/PhysRevC.110.064320}
}

@article{Zhang_Lacroix_2025,
  author  = {Zhang, J. and Lacroix, D.},
  title   = {Excited States from {ADAPT-VQE} Convergence Path in Many-Body Problems: Application to Nuclear Pairing Problem and {$H_4$} Molecule Dissociation},
  journal = {Phys. Lett. B},
  volume  = {869},
  pages   = {139841},
  year    = {2025},
  doi     = {10.1016/j.physletb.2025.139841}
}

@article{Yoshida_2026,
  author  = {Yoshida, S. and Sato, T. and Ogata, T. and Kimura, M.},
  title   = {Bridging Quantum Computing and Nuclear Structure: Atomic Nuclei on a Trapped-Ion Quantum Computer},
  journal = {Phys. Rev. Res.},
  volume  = {8},
  pages   = {013134},
  year    = {2026},
  doi     = {10.1103/td9s-z7my}
}

@misc{MataAli_2025,
  author        = {Mata Ali, A.},
  title         = {Fast Expectation Value Calculation Speedup of Quantum Approximate Optimization Algorithm: {HoLCUs QAOA}},
  year          = {2025},
  eprint        = {2503.01748},
  archivePrefix = {arXiv},
  primaryClass  = {quant-ph}
}

@article{Childs_Wiebe_2012,
  author  = {Childs, A. M. and Wiebe, N.},
  title   = {Hamiltonian Simulation Using Linear Combinations of Unitary Operations},
  journal = {Quantum Inf. Comput.},
  volume  = {12},
  number  = {11--12},
  pages   = {901--924},
  year    = {2012},
  doi     = {10.26421/qic12.11-12-1}
}

@article{Berry_2015,
  author  = {Berry, D. W. and Childs, A. M. and Cleve, R. and Kothari, R. and Somma, R. D.},
  title   = {Simulating Hamiltonian Dynamics with a Truncated Taylor Series},
  journal = {Phys. Rev. Lett.},
  volume  = {114},
  number  = {9},
  pages   = {090502},
  year    = {2015},
  doi     = {10.1103/PhysRevLett.114.090502}
}

@misc{Grover_Rudolph_2002,
  author        = {Grover, L. and Rudolph, T.},
  title         = {Creating Superpositions That Correspond to Efficiently Integrable Probability Distributions},
  year          = {2002},
  eprint        = {quant-ph/0208112},
  archivePrefix = {arXiv},
  primaryClass  = {quant-ph}
}

@article{Ramacciotti_2024,
  author  = {Ramacciotti, D. and Lefterovici, A. I. and Rotundo, A. F.},
  title   = {Simple Quantum Algorithm to Efficiently Prepare Sparse States},
  journal = {Phys. Rev. A},
  volume  = {110},
  number  = {3},
  pages   = {032609},
  year    = {2024},
  doi     = {10.1103/PhysRevA.110.032609}
}

@article{Mottonen_2004,
  author  = {M{\"o}tt{\"o}nen, M. and Vartiainen, J. J. and Bergholm, V. and Salomaa, M. M.},
  title   = {Quantum circuits for general multiqubit gates},
  journal = {Phys. Rev. Lett.},
  volume  = {93},
  pages   = {130502},
  year    = {2004},
  doi     = {10.1103/PhysRevLett.93.130502}
}

@misc{Sun_2023,
  author        = {Sun, Xiaoming and Tian, Guojing and Yang, Shuai and Yuan, Pei and Zhang, Shengyu},
  title         = {Asymptotically Optimal Circuit Depth for Quantum State Preparation and General Unitary Synthesis},
  year          = {2023},
  eprint        = {2108.06150},
  archivePrefix = {arXiv},
  primaryClass  = {quant-ph}
}

@misc{Jaunin_2025,
  author       = {Jaunin, H.},
  title        = {\href{https://pennylane.ai/blog/2025/10/state-preparation/}{How Quantum State Preparation Can Make or Break Your Hackathon Project}},
  year         = {2025},
  month        = {October},
  howpublished = {PennyLane Blog}
}

@article{Babbush_2018,
  author  = {Babbush, Ryan and Gidney, Craig and Berry, Dominic W. and Wiebe, Nathan and McClean, Jarrod and Paler, Alexandru and Fowler, Austin and Neven, Hartmut},
  title   = {Encoding Electronic Spectra in Quantum Circuits with Linear T Complexity},
  journal = {Phys. Rev. X},
  volume  = {8},
  number  = {4},
  pages   = {041015},
  year    = {2018},
  doi     = {10.1103/PhysRevX.8.041015}
}

@misc{Wierichs_2025,
  author       = {Wierichs, D.},
  title        = {\href{https://pennylane.ai/compilation/unary-iteration}{Unary Iteration}},
  year         = {2025},
  howpublished = {PennyLane Quantum Compilation}
}

@book{Suhonen_2007,
  author    = {Suhonen, J.},
  title     = {From Nucleons to Nucleus: Concepts of Microscopic Nuclear Theory},
  publisher = {Springer},
  address   = {Berlin},
  year      = {2007}
}

@misc{Johnson_2018,
  author        = {Johnson, C. W. and Ormand, W. E. and McElvain, K. S. and Shan, H.},
  title         = {{BIGSTICK}: A Flexible Configuration-Interaction Shell-Model Code},
  year          = {2018},
  eprint        = {1801.08432},
  archivePrefix = {arXiv},
  primaryClass  = {nucl-th}
}

@article{Cohen_Kurath_1965,
  author  = {Cohen, S. and Kurath, D.},
  title   = {Effective Interactions for the {$1p$} Shell},
  journal = {Nucl. Phys.},
  volume  = {73},
  number  = {1},
  pages   = {1--24},
  year    = {1965},
  doi     = {10.1016/0029-5582(65)90148-3}
}

@article{Richter_2008,
  author  = {Richter, W. A. and Mkhize, S. and Brown, B. A.},
  title   = {{$\mathit{sd}$}-Shell Observables for the {USDA} and {USDB} Hamiltonians},
  journal = {Phys. Rev. C},
  volume  = {78},
  number  = {6},
  pages   = {064302},
  year    = {2008},
  doi     = {10.1103/PhysRevC.78.064302}
}

@misc{JavadiAbhari_2024,
  author        = {Javadi-Abhari, A. and Treinish, M. and Krsulich, K. and Wood, C. J. and Lishman, J. and Gacon, J. and Martiel, S. and Nation, P. D. and Bishop, L. S. and Cross, A. W. and Johnson, B. R. and Gambetta, J. M.},
  title         = {Quantum Computing with {Qiskit}},
  year          = {2024},
  eprint        = {2405.08810},
  archivePrefix = {arXiv},
  primaryClass  = {quant-ph}
}

@article{Gidney_2018,
  author  = {Gidney, C.},
  title   = {Halving the Cost of Quantum Addition},
  journal = {Quantum},
  volume  = {2},
  pages   = {74},
  year    = {2018},
  doi     = {10.22331/q-2018-06-18-74}
}

\end{document}